\pdfoutput=1
\documentclass[11pt, a4paper]{article}
\usepackage{jcappub.local}

\makeatletter
\@addtoreset{equation}{section}
\makeatother

\allowdisplaybreaks

\def\bk{{\bf k}}

\def\CH{{\cal H}}
\def\CL{{\cal L}}

\def\ttau{{\tilde \tau}}

\def\half{\frac{1}{2}}

\usepackage{subfigure}
\usepackage[subfigure]{graphfig}

\usepackage[section]{placeins}


\title{Strongly Coupled Quasi-Single Field Inflation}

\author[a]{Aditya Varna Iyer}
\author[b]{Shi Pi}
\author[a,c]{Yi Wang}
\author[d,e]{Ziwei Wang}
\author[a,c]{Siyi Zhou}

\affiliation[a]{Department of Physics, The Hong Kong University of Science and Technology,\\
Clear Water Bay, Kowloon, Hong Kong, P.R.China}
\affiliation[b]{Center for Gravitational Physics, Yukawa Institute for Theoretical Physics, \\
Kyoto University, Kyoto 606-8502, Japan}
\affiliation[c]{Jockey Club Institute for Advanced Study, The Hong Kong University of Science and Technology, \\
Clear Water Bay, Kowloon, Hong Kong, P.R.China}
\affiliation[d]{Physics Department, McGill University, Montreal, QC, H3A 2T8, Canada}
\affiliation[e]{School of Physics, University of Science and Technology of China, Hefei, Anhui, 230026,
P.R.China}

\abstract{We study the power spectrum of quasi-single field inflation where strong coupling is considered. The contribution from the massive propagator can be divided into local and non-local contributions. The local one is the leading contribution and is power-law suppressed as a function of mass, while the non-local contribution is exponentially suppressed in the large mass limit. For the local contribution, it is possible to use the effective field theory approach to study the power spectrum in the strongly coupled region of the parameter space. For the non-local contribution, we developed a \textit{partial effective field theory} method to simplify the calculation: When there are multiple massive propagators, one can fully compute it after integrating out all but one massive propagator by effective field theory. The result retains the ``standard clock'' signal, which is interesting for probing the expansion history of the primordial universe and the physics of a ``cosmological collider''. The error involved compared to the full calculation is power law suppressed by the effective mass of the heavy field.}

\begin{document}
\begin{flushright}
    YITP-17-105
\end{flushright}
\maketitle

\section{Introduction} \label{sec:intro}
An interesting question pertaining to the primordial universe is to detect the particle content during inflation. In all UV complete models of inflation, there exist additional massive fields \cite{Linde:1993cn,Yamaguchi:2005qm,Chen:2009we, Chen:2009zp,Tolley:2009fg,Achucarro:2010jv,Cremonini:2010ua}. If those additional fields have masses much smaller than the Hubble parameter $H$, they could as well drive inflation as the inflaton, connoting multi-field inflation~\cite{Sasaki:1998ug,Gordon:2000hv,Amendola:2001ni,Peterson:2010np}. If the additional fields have masses much larger than $H$, they can be systematically integrated out of the framework thus leaving an effective single field inflation model. The remaining fields of mass $m\sim H$ are of particular interest because they can leave distinct non-Gaussianities, which has been studied in a typical model known as the quasi-single field inflation \cite{Chen:2009we, Chen:2009zp}.

Such non-Gaussianities are particularly useful in several aspects. For instance, by looking at the squeezed limit of the non-Gaussianity, we can extrapolate information about the mass \cite{Chen:2009we, Chen:2009zp, Arkani-Hamed:2015bza} and spin \cite{Arkani-Hamed:2015bza} of the massive particle present during inflation. This is later generalized to an arbitrary triangle momentum configuration in \cite{Lee:2016vti}. The spectrum of the massive particles due to the standard model uplifting is studied in \cite{Chen:2016uwp}. There are of course other possible playgrounds for massive particles in the primordial universe. They can serve as a standard clock to model-independently distinguish the early universe scenarios \cite{Chen:2015lza,Chen:2016cbe,Chen:2016qce,Huang:2016quc}.

This distinctive signature is however, quite small in amplitude, in general. One reason for this is by virtue of the Boltzmann suppression $\sim e^{-\pi\mu}$ with $\mu\sim \frac{m}{H}$ \cite{Gibbons:1977mu,Mottola:1984ar,Ford:1986sy}. The massive fields can be thought of as massive particles in the early universe. Their statistics satisfies the Boltzmann distribution, leading to significant suppression. The other reason stems from the fact that if one considers all the possible couplings of the quasi-single field inflation, it is found that the three-point interactions of the massive field and the primordial curvature perturbation are all at least suppressed by slow-roll parameters except for the diagram that gains contributions from the self-interaction of the massive field $\delta\sigma^3$ \cite{Chen:2009zp}. Followed by three two-point transfer vertices playing the role of connecting the massive field with the curvature perturbation, this type of contribution can evade the slow-roll suppression. Actually, this three-heavy-field interaction can possibly produce a large $f_{\rm NL}$ which may hopefully be observable in future analyses of the CMB large scale structure, and the 21 cm surveys \cite{Meerburg:2016zdz}.

An unsatisfactory aspect of this interaction is that the analytical result is not yet known due to the difficulty involved in integrating multiple products of the Hankel functions in the in-in formalism, except in the large mass limit \cite{Gong:2013sma}. 
The bispectrum involving contributions from the $\delta\sigma\dot{\delta\theta}\dot{\delta\theta}$ interaction, however, is easier to compute. One may naturally think about combining the merits of these two kinds of interactions. The idea is that every massive field contains two contributions, the vacuum type contribution and the thermal type contribution. The vacuum contribution is the most dominant one. If we are on only interested in the first order correction to the vacuum contribution, the three-point-function with $\delta\sigma^3$ interaction can be simplified as an effective theory of $\delta\sigma\dot{\delta\theta}\dot{\delta\theta}$ interaction. The realization is to integrate out only two of the massive fields, but leave one massive field unintegrated. This can preserve the leading-order nonlocal effects from the heavy field, at the same time keep the possibility for analytical calculation. We dubbed this method as \textit{partial effective field theory}.

The effective field theory approach is very useful to study the inflationary models with the presence of one or more massive fields~\cite{Weinberg:2008hq}. 
It has been shown that when the mass hierarchies are large, those massive fields can be integrated out, thus yielding a single field inflation model with a modified sound speed $c_s^{-2} = 1+4 \dot\theta_0^2 / m_{\rm eff}^2$, where $\dot\theta_0$ is the velocity of the inflaton background and the $m_{\rm eff}$ is the effective mass of the additional massive field \cite{Tolley:2009fg,Gwyn:2014doa,Achucarro:2010da,Achucarro:2012sm,Chen:2012ge,Pi:2012gf}. A discussion focusing on the symmetry consideration appears in \cite{Mirbabayi:2015hva}. For instance, to calculate the non-Gaussianity induced by the three-heavy-field vertex, we can integrate all the heavy fields out, and get its leading-order local contribution in the three-point function of the curvature perturbation~\cite{Gong:2013sma}. 
However, in this paper, as we are also interested in the nonlocal effect of the massive field, we integrate out only part of the massive field(s) contribution while keeping one massive degree of freedom, which in turn will give us the leading-order nonlocal effects in the large mass limit.

The best studied model of quasi-single field inflation is that with an inflaton and a heavy field moving in a constant-turn trajectory~\cite{Chen:2009we,Chen:2009zp}, especially in the weakly coupled regime where $\dot{\theta_0}<m_{\rm eff}$ \cite{Chen:2012ge,Pi:2012gf,Noumi:2012vr}. In this regime, the usual perturbative method: in-in formalism can be applied. The discussion can be extended to strongly coupled regime $\dot{\theta_0}> m_{\rm eff}$ \cite{Baumann:2011su,Cremonini:2010ua,Gwyn:2012mw,An:2017hlx,Tong:2017iat} using effective field theory method. One thing we should notice is that there are two types of strongly coupled regime, the large mass strongly coupled regime $(m_\text{eff}<\dot\theta_0<m_\text{eff}^2/H)$ and the extremely strongly coupled regime $(H\dot\theta_0>m_\text{eff}^2)$. The effective field theory of both strongly coupled regimes is obtained in the papers \cite{Baumann:2011su,Gwyn:2012mw}, while the power spectrum of both are studied in \cite{Cremonini:2010ua} by solving the coupled differential equation recursively. The result in the large mass strongly coupled regime is also applicable to the case of weakly coupled regime. A uniform expression applicable to all the mass and coupling satisfying $m_{\rm eff}^2+\dot\theta_0^2>9H^2/4 $ using horizon crossing approximation is obtained in \cite{Tong:2017iat}, which is in agreement with the result in two regimes \cite{Cremonini:2010ua} and the numerical result in \cite{An:2017hlx}. For the bispectrum, \cite{An:2017hlx} studied the analytical part contributed by virtual particle production analytically. The scaling behavior of non-analytical clock signal is extracted to be $(k_1/k_3)^{i\sqrt{m_{\rm eff}^2/H^2+{\dot \theta_0^2/H^2}}}$ using IR analysis of the differential equation. The prefactor of the clock signal is obtained by fitting the IR behavior of the mode function and solving the coupled differential equations numerically.

In this work, we are going to discuss the full parameter space for the coupling as long as it satisfies $H\dot\theta_0/m_\text{eff}^2<1$~\cite{Tong:2017iat}. We use the effective field theory method to obtain the power spectrum in this regime. We are able to verify this result by using the Schwinger-Keldysh diagrammatics recently developed in \cite{Chen:2017ryl} up to an arbitrary order. When the system is strongly coupled, it is enough to replace the original mode function by the effective mode function with a modification of the sound speed. The same method applies to the bispectrum. We are able to study the bispectrum systematically by applying both the partial effective field theory and the effective mode function.

This paper is organized as follows. In Sec. \ref{QSFI}, we review the constant-turn quasi-single field inflation.  In Sec. \ref{EFTforBispectrum} we apply the partial effective field theory approach to quasi-single field inflation with large mass, which is shown to yield the leading-order nonlocal effects in the large mass limit. 
In Sec. \ref{Strongly}, three methods, i.e. the numerical solution of the equation of motion, the partial effective field theory, and the Schwinger-Keldysh diagrammatics, are applied to extend the previous study to the strong coupling regime. In Sec. \ref{correctedbispectrum}, we proceed to shed light on the bispectrum using the method introduced.

\section{Quasi-Single Field Inflation}\label{QSFI}
\setcounter{equation}{0}
The purpose of this section is to put forth a model of quasi-single field inflation akin to the ``constant turn" case introduced in \cite{Chen:2009we,Chen:2009zp}. The model chosen incorporates the idea of decomposing the light fields along the flat slow-roll direction while the massive ones are decomposed along the mutually perpendicular isocurvature direction. The action is written in terms of the radius $\tilde R$ for the arc of the circular local minima in field space with $\tilde R\theta$ and $\sigma$ being the fields on the tangential and radial directions of the trajectory as follows,
\begin{align}
 S_m = \int d^4x \sqrt{-g} \left[
 -\half (\tilde R+\sigma)^2 g^{\mu\nu} \partial_\mu \theta \partial_\nu \theta
 - \half g^{\mu \nu} \partial_\mu \sigma \partial_\nu \sigma
 - V_{\rm sr}(\theta) - V(\sigma) \right] ~,
 \label{ModelAction}
\end{align}
The separated potential comprises of the conventional slow-roll potential $V_{\rm sr}(\theta)$ and the perpendicular potential $V(\sigma)$ influencing the isocuravton. The usual equations of motion for the case similar to Single field inflation are
\begin{align}
   & 3M_p^2 H^2 =\frac{1}{2}R^2\dot\theta_0^2+V (\sigma) +V_{\rm sr} (\theta)~, \\
   & -2M_p^2 \dot H = R^2\dot\theta_0^2 ~,                                      \\
   & \sigma_0 = {\rm const.} \quad,\quad V'(\sigma_0) = R\dot{\theta}_0^2       \\
   & R^2 \ddot{\theta}_0 + 3 R^2 H \dot{\theta}_0 + V'_{\rm sr} = 0~.
\end{align}
where the $\sigma_0(t)$ and $\theta_0(t)$ are the backgrounds of the massive field and the inflaton, respectively, and $R=\tilde R+\sigma_0$ is the radius of the background trajectory in the field space. By virtue of $V(\sigma)$, the isocuravton tends to be trapped at $\sigma_0$, We choose to expand the potential around this point as
\begin{align}
 V(\sigma) = V'(\sigma_0)  (\sigma-\sigma_0) + \half V''(\sigma_0) (\sigma-\sigma_0)^2 + \frac{1}{6} V'''(\sigma_0) (\sigma-\sigma_0)^3 + \cdots~.
\end{align}
As pointed out in \cite{Chen:2009zp}, the interaction that is likely to produce large non-Gaussianity is this three-point self-interaction of the massive field. 
However, after having performed this expansion we stipulate that the terms involving the derivatives of the potential be written simply as $V', V''$ and $V'''$ while realizing that they depend on $\sigma_0$ for convenience.  We choose the spatially flat gauge and introduce a small perturbation to the inflaton and the isocurvaton,
\begin{align}
 \theta(\boldsymbol x,t) = \theta_0(t) + \delta\theta(\boldsymbol x,t) ~, ~~~~~
 \sigma(\boldsymbol x,t) = \sigma_0 + \delta\sigma(\boldsymbol x,t) ~.
 \label{gauge1fields}
\end{align}
Plugging these into the Lagrangian while ignoring the gravity part enables us to obtain the following
\begin{align}
 \CL_2 = \frac{a^3}{2} R^2 \dot {\delta\theta}^2 - \frac{a R^2}{2}
 (\partial_i \delta\theta)^2
 + \frac{a^3}{2} \dot{\delta\sigma}^2 - \frac{a}{2} (\partial_i
 \delta\sigma)^2 -\frac{a^3}{2} (V''-\dot\theta_0^2) \delta \sigma^2
 ~,
 \label{CL2}
\end{align}
\begin{align}
 \delta \CL_2 = & 2 a^3 R \dot \theta_0 ~\delta\sigma \dot{\delta\theta} = c_2 a^3 \delta\sigma \dot{\delta\theta}
 ~,
 \label{dCL2} \\
 \CL_3 =        & -\frac{1}{6} a^3 V''' \delta\sigma^3  = - c_3a^3\delta\sigma^3~.
 \label{CL3}
\end{align}
Here $\CL_2$ describes the dynamics of the $\sigma$ and $\theta$ fields while the coupling between them is given by $\delta \CL_2$ and $\CL_3$ is the leading source for the three-point function involving $\delta\sigma$.

To obtain the Hamiltonian density it is necessary to define the conjugate momenta following which we separate the obtained result into  the free part $\CH_0$, and the interacting part $\CH_I$.  The in-in formalism can be used if we replace the conjugate momenta in the Hamiltonian density with those corresponding to the interacting picture. Such a procedure gives
\begin{align} \label{H0}
 \CH_0 = a^3 \left[ \half R^2 \dot {\delta\theta_I}^2 +
 \frac{R^2}{2a^2}
 (\partial_i \delta\theta_I)^2
 + \half \dot{\delta\sigma_I}^2 + \frac{1}{2a^2} (\partial_i
 \delta\sigma_I)^2 + \half m_\text{eff}^2 \delta \sigma_I^2
 \right] ~,
\end{align}
and interaction Hamiltonian density
\begin{align}
 \CH^I_2 = & -c_2 a^3 \delta\sigma_I \dot{\delta\theta_I} ~,
 \label{CH2}
 \\
 \CH^I_3 = & c_3 a^3 \delta\sigma_I^3 ~,
 \label{CH3}
\end{align}
where $m_\text{eff}\equiv V''-\dot\theta^2$ is the effective mass of the heavy field. We proceed to decompose the interacting fields in terms of their mode functions as follows
\begin{align}
 \delta\theta_{\boldsymbol k}^I = u_{\boldsymbol k} a_{\boldsymbol k} + u_{-\boldsymbol k}^* a_{-\boldsymbol k}^\dagger ~,
                                                                                                                           \\
 \delta\sigma_\bk^I  =  v_{\boldsymbol k} b_{\boldsymbol k} + v_{-{\boldsymbol k}}^* b_{-\boldsymbol k}^\dagger ~,
\end{align}
The usual commutation relations are obeyed by the creation and annihilation operators,
\begin{align}
 [a_{\boldsymbol k},a_{-{\boldsymbol k}'}^\dagger] = (2\pi)^3 \delta^3 ({\boldsymbol k}+{\boldsymbol k}') ~,
 \quad
 [b_{\boldsymbol k},b_{-{\boldsymbol k}'}^\dagger] = (2\pi)^3 \delta^3 ({\boldsymbol k}+{\boldsymbol k}') ~.
\end{align}
while the mode functions themselves follow the equations of motion below
\begin{align}
 u_{\boldsymbol k}'' - \frac{2}{\tau} u_{\boldsymbol k}' + k^2 u_{\boldsymbol k} =0 ~,
 \label{modefun_u}                                                                     \\
 v_{\boldsymbol k}'' - \frac{2}{\tau} v_{\boldsymbol k}' + k^2 v_{\boldsymbol k} +
 \frac{m^2}{H^2 \tau^2} v_{\boldsymbol k} = 0 ~,
 \label{modefun_v}
\end{align}
The prime denotes derivative with respect to the conformal time $\tau$. The equations of motion  for $u_\textbf{k}$ can be solved in a usual way,
\begin{align}
 u_{\boldsymbol k} = \frac{H}{R\sqrt{2k^3}} ( 1+i k\tau)e^{-i k\tau} ~,
 \label{mode_u}
\end{align}
while the mode function for $v_{\boldsymbol k}$ will be expressed by the Hankel function of the first kind,
\begin{align}
 v_{\boldsymbol k} = -i e^{ -\frac{\pi}{2}\mu + i\frac{\pi}{4}}
 \frac{\sqrt{\pi}}{2} H
 (-\tau)^{3/2} H^{(1)}_{i\mu} (-k\tau) ~,
 \label{mode_v2}
\end{align}
with $\mu = \sqrt{\frac{m^2}{H^2}-\frac{9}{4}}$. Note that the order of the Hankel function is pure imaginary for large mass: $m^2/H^2 >9/4$.

\begin{figure}[htbp]
 \centering
 \includegraphics[width=0.2\textwidth]{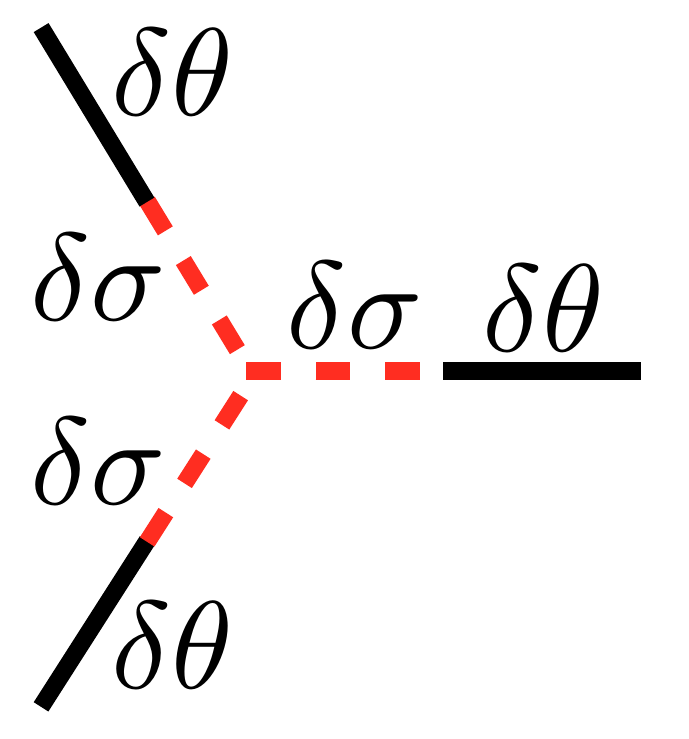}
 \caption{\label{fig:shape} This is the diagram associated with the self-interaction of the massive field. We can see that this Feynman diagram consists of one self-interaction of the massive field $\delta\sigma^3$ and three transfer vertex $\delta\sigma\dot{\delta\theta}$. This is the only diagram in the quasi-single field inflation model which can produce large non-Gaussianities. }
 \label{3sigmadiagram}
\end{figure}

Next we try to calculate the bispectrum induced by the  cubic self-interaction of the massive field, which is shown schematically in Figure~\ref{3sigmadiagram}. This term may give rise to observable non-Gaussianities. The interaction Hamiltonian is obtained by integrating the term in (2.11) over space which is the leading cubic term necessary for the bispectra calculations,
\begin{align}
 H^I_3 =\int d^3\boldsymbol x \CH^I_3
 = c_3 a^3 \int \frac{d^3{\boldsymbol p}}{(2\pi)^3} \frac{d^3{\boldsymbol q}}{(2\pi)^3}
 \delta\sigma^I_{\boldsymbol p}(t) \delta\sigma^I_{\boldsymbol q}(t)
 \delta\sigma^I_{-{\boldsymbol p}-{\boldsymbol q}}(t) ~.
\end{align}
The transfer vertex is
\begin{align}
 H^I_2 = \int d^3\boldsymbol x \mathcal H_2^I = - c_2 a^3 \int \frac{d^3 \boldsymbol p}{(2\pi)^3} \delta\sigma^I_{\boldsymbol p} \dot{\delta\theta}^I_{-\boldsymbol p}~.
\end{align}
It is introduced to convert the isocurvature mode into the curvature mode hence expressed as non-Gaussianities of the curvature perturbation.

We use the in-in formalism to write the three-point function as follows
\begin{align}
 \langle \delta\theta^3 \rangle \equiv
 \langle 0| \left[ \bar T \exp\left( i\int_{-\infty}^0 d\tilde\tau H_I(\tilde\tau)\right)
 \right] \delta\theta_I^3(0)
 \left[ T \exp\left( -i\int_{-\infty}^0 d\tau H_I(\tau)\right)
 \right] |0\rangle ~.
 \label{3ptinin}
\end{align}
In the following lines, we discuss the ``factorized form'' of the in-in formalism, which contains the fourth order Hamiltonian as
\begin{align}
   & \langle \delta\theta^3 \rangle
 = \int_{-\infty}^0 d\tilde \tau_1 \int_{-\infty}^{\tilde \tau_1}
 d\tilde \tau_2
 \int_{-\infty}^0 d\tau_1 \int_{-\infty}^{\tau_1} d\tau_2
 ~\langle H_I(\tilde \tau_2) H_I(\tilde \tau_1) ~\delta\theta_I^3~
 H_I(\tau_1)H_I(\tau_2) \rangle
 \label{3pt1} \\
   & - 2 ~{\rm Re} \left[
 \int_{-\infty}^{0}d\tilde \tau_1
 \int_{-\infty}^0 d\tau_1 \int_{-\infty}^{\tau_1} d\tau_2 \int_{-\infty}^{\tau_2} d\tau_3
 ~\langle H_I(\tilde \tau_1) ~\delta\theta_I^3~
 H_I(\tau_1) H_I(\tau_2) H_I(\tau_3) \rangle
 \right]
 \label{3pt2}\\
   & + 2 ~{\rm Re} \left[
 \int_{-\infty}^0 d\tau_1 \int_{-\infty}^{\tau_1} d\tau_2 \int_{-\infty}^{\tau_2} d\tau_3
 \int_{-\infty}^{\tau_3} d\tau_4
 ~\langle \delta\theta_I^3~
 H_I(\tau_1) H_I(\tau_2) H_I(\tau_3) H_I(\tau_4) \rangle
 \right] ~.
 \label{3pt3}
\end{align}
We can see from Figure.\ref{3sigmadiagram} that the transfer vertex $H^I_2$  appears in any three of the four terms labelled $H_I$ while the remaining term corresponds to the interaction vertex $H^I_3$.  This gives rise to the following terms after contractions,
\begin{align}
     & - 12 c_2^3 c_3 u_{k_1}^*(0) u_{k_2}(0) u_{k_3}(0)
 \cr & \times
 {\rm Re} \left[ \int_{-\infty}^0 d\ttau_1~ a^3(\ttau_1)
 v_{k_1}^*(\ttau_1) u'_{k_1}(\ttau_1)
 \int_{-\infty}^{\ttau_1} d\ttau_2~ a^4(\ttau_2) v_{k_1}(\ttau_2)
 v_{k_2}(\ttau_2) v_{k_3}(\ttau_2) \right.
 \cr & \times \left.
 \int_{-\infty}^0 d\tau_1~ a^3(\tau_1) v_{k_2}^*(\tau_1)
 u_{k_2}^{\prime *} (\tau_1)
 \int_{-\infty}^{\tau_1} d\tau_2~ a^3(\tau_2) v_{k_3}^*(\tau_2)
 u_{k_3}^{\prime *} (\tau_2) \right]
 \cr & \times
 (2\pi)^3 \delta^3( \boldsymbol k_1+ \boldsymbol k_2+ \boldsymbol k_3) + {\rm 9~other~similar~terms}
 \cr
     & + {\rm 5~permutations~of~} \boldsymbol k_i ~.
 \label{FacForm_term}
\end{align}
The detailed result of these 10 terms with the mode functions \eqref{mode_u} and \eqref{mode_v2} are presented in Appendix \ref{concreteform}.

\section{Partial Effective Field Theory Approach to Quasi-Single Field Inflation}\label{EFTforBispectrum}
\setcounter{equation}{0}
In this section, we would like to introduce a new kind of effective field theory approach to deal with the quasi-single field inflation with large mass. We will focus on the three-point function  with the $\delta\sigma^3$ interaction described in the proceeding section.

\begin{figure}[htbp]
 \centering
 \includegraphics[width=0.2\textwidth]{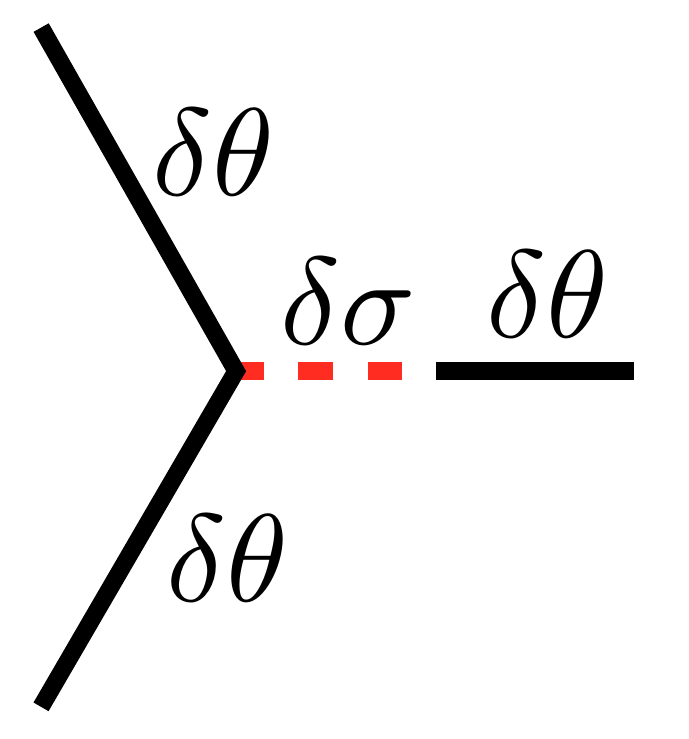}
 \caption{\label{fig:shape} This is the three-point function with the $\delta\sigma\dot{\delta\theta}\dot{\delta\theta}$ interaction. Since it only has two interacting vertices, the three-point correlation has a simpler form than that contributed by the $\delta\sigma\delta\sigma\delta\sigma$ interaction.}
 \label{1sigmadiagram}
\end{figure}

Two varieties of methods are used to systematically show that, if we are interested in the result only up to the $\frac{1}{\mu^4} e^{-\pi\mu}$ order, we can simply compute three-point function shown in Figure \ref{1sigmadiagram}.  This method can be regarded as a generalization of the method introduced in \cite{Gong:2013sma}, where two kinds of methods are used to prove that 
the quasi-single field inflation effectively reduces to single field inflation with a modification of the sound speed and the interacting vertex in the large mass limit.  One way to show this is to integrate out all the three heavy fields, degenerate its equation of motion to be an algebraic relation, and substitute its solution back to the action to obtain an effective single field Lagrangian with a modified sound speed and a cubic interaction term of the curvature perturbation. The second method is the large mass limit of in-in formulism, in which the large mass approximation of the Hankel function in \eqref{mode_v2} is used for the $\delta\sigma$ propagator. The results from these two methods are the same up to the leading order.

The advantage of these methods used in \cite{Gong:2013sma} is to give an analytical expression for the three-point function induced by the three-heavy-field vertex shown in Figure~\ref{3sigmadiagram}, while its disadvantage is that all the information on the nonlocal effects, which will turn out to be the oscillation features in the bispectrum of the curvature perturbation, is completely neglected. Therefore, instead of adopting this method for all three of the massive $\delta\sigma$ propagators, in our partial effective field theory approach, we would like to use it for only two of the three massive $\delta\sigma$ propagators. More concretely, we examine the squeezed limit $k_1 = k_2 \gg k_3$. This would mean that the mode functions corresponding to momenta $k_1$ and $k_2$ are short wavelength modes and that they are more likely to exhibit a vacuum contribution while $k_3$ is the long wavelength mode and more likely to exhibit a thermal contribution. Hence for the partial EFT approach, we keep the massive field with the $k_3$ momentum unchanged while treating the others as effective field after integrated out. It turns out that this new partial effective field theory approach is successful and a good approximation up to the order $\frac{1}{\mu^4}e^{-\pi\mu}$. The calculation of the in-in expectation value of Figure \ref{3sigmadiagram} can be successfully approximated to be the in-in expectation value of Figure \ref{1sigmadiagram}, while in the latter one the leading oscillatory features from the massive field are preserved.

\subsection{Order Estimation}\label{order}
\begin{figure}[htbp]
 \centering
 \includegraphics[width=0.4\textwidth]{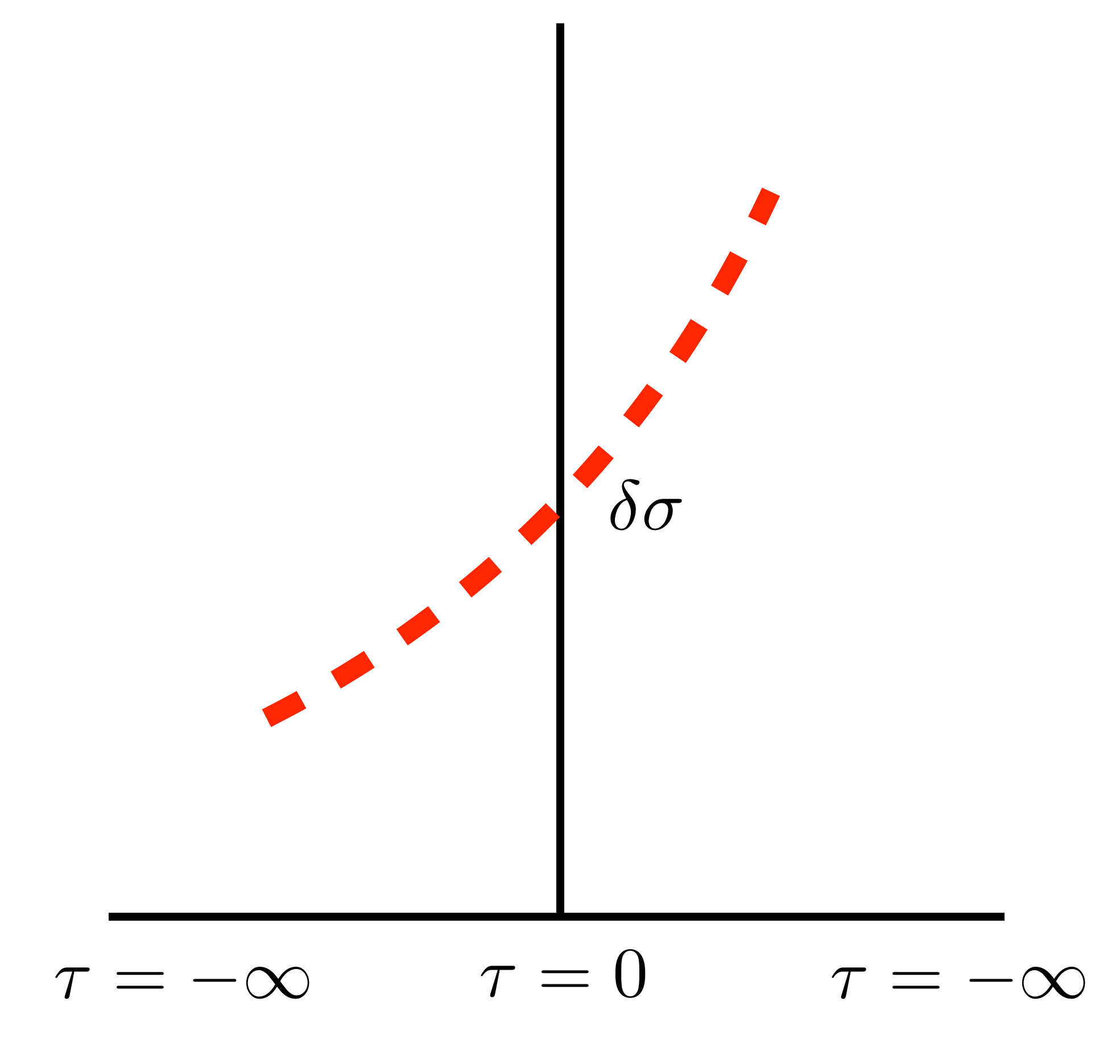} \quad
 \includegraphics[width=0.4\textwidth]{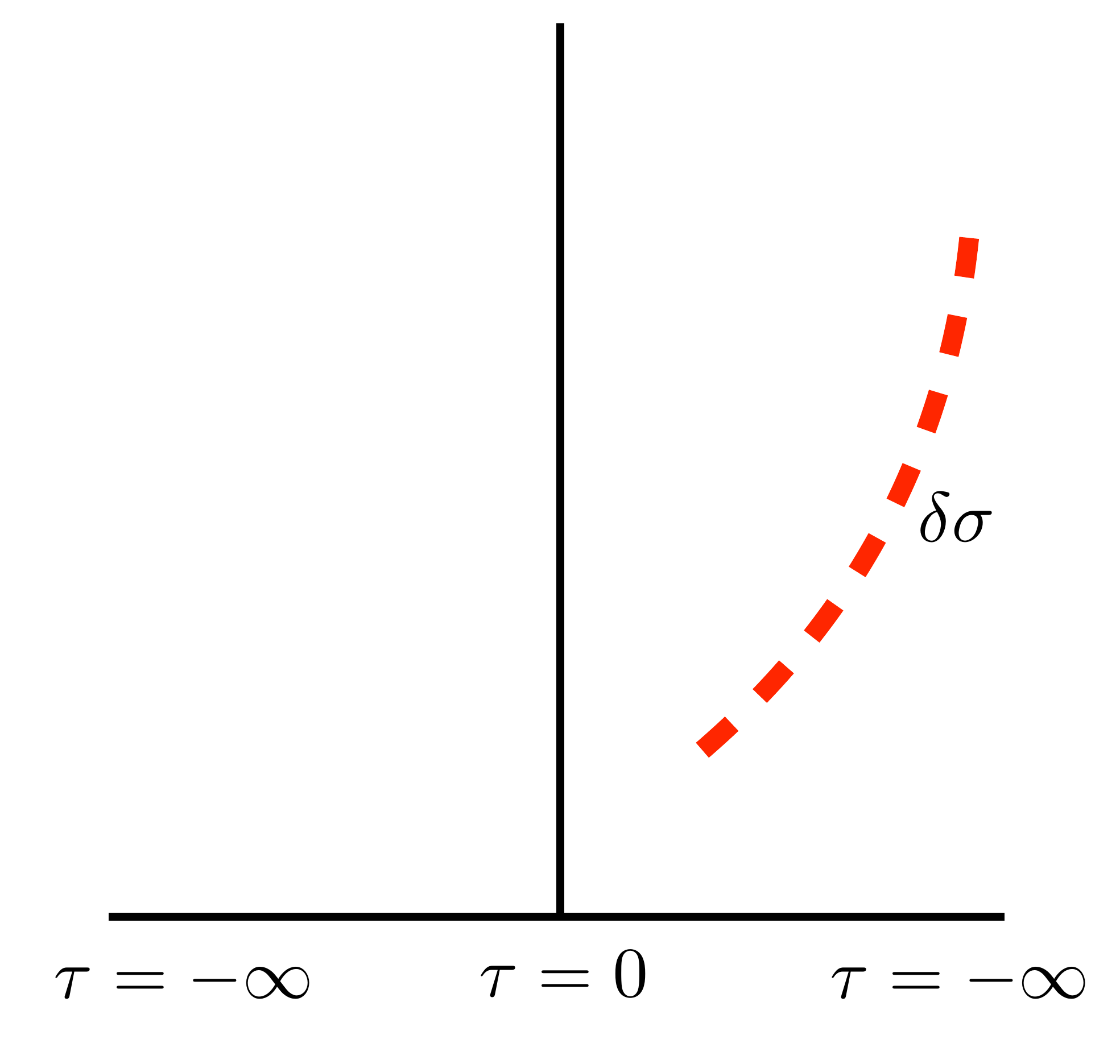}
 \caption{\label{fig:shape} These two plots are two possible $\delta\sigma$ propagators that may contribute to the diagram in Figure \ref{3sigmadiagram}. The vertical lines stand for the $\tau = 0$ time surface. The left hand side of the vertical lines denotes the non-time-ordered contribution $D_{+-(-+)}$. The right hand side denotes the time ordered contribution $D_{++(--)}$.}
 \label{Dpropagators}
\end{figure}

Before proceed we introduce some order estimation for the 10 integrals given in \eqref{FacForm_term} and in Appendix~\ref{concreteform}.
It is observed that there are two possible $\langle\delta\sigma(\tau_1)\delta\sigma(\tau_2)\rangle$ propagators. One is the non-time-ordered propagator, in the sense that $\tau_1$ and $\tau_2$ varies freely. The other one is the time ordered $\delta\sigma$ propagator, where $\tau_1$ should be prior than $\tau_2$ or vice versa. This can be seen more explicitly from the Schwinger-Keldish formalism, where the propagators can be written in terms of the form
\begin{align}
 G_{++} (k;\tau_1,\tau_2) & = u_{k} (\tau_1) u^*_{k} (\tau_2) \theta (\tau_1- \tau_2) + u^*_k (\tau_1) u_k(\tau_2) \theta (\tau_2 - \tau_1),    \\
 G_{+-} (k;\tau_1,\tau_2) & =  u^*_k (\tau_1) u_k(\tau_2),                                                                                      \\
 G_{-+} (k;\tau_1,\tau_2) & =  u_{k} (\tau_1) u^*_{k} (\tau_2),                                                                                 \\
 G_{--} (k;\tau_1,\tau_2) & = u^*_k (\tau_1) u_k(\tau_2) \theta  (\tau_1- \tau_2) +  u_{k} (\tau_1) u^*_{k} (\tau_2)  \theta (\tau_2 - \tau_1),
\end{align}
for the massless field and
\begin{align}
 D_{++}(\tau_1,\tau_2) = & v_k(\tau_1)v_k^{*}(\tau_2) \Theta(\tau_1 - \tau_2) +
 v_k^*(\tau_1)v_k(\tau_2) \Theta(\tau_2-\tau_1),  \\
 D_{+-}(\tau_1,\tau_2) = & v_k^*(\tau_1)v_k(\tau_2),                            \\
 D_{-+}(\tau_1,\tau_2) = & v_k(\tau_1)v_k^{*}(\tau_2),                          \\
 D_{--}(\tau_1,\tau_2) = & v_k^*(\tau_1)v_k(\tau_2)\Theta(\tau_1-\tau_2)+
 v_k(\tau_1)v_k^{*}(\tau_2) \Theta(\tau_2-\tau_1).
\end{align}
for the massive field. Here we want to conduct an order estimation for the massive propagators, while the massless propagators are introduced for future convenience. $D_{++}$ and $D_{--}$ types of propagators can be drawn diagramatically as the left hand side of Figure~\ref{Dpropagators}, whereas the mixed propagators $D_{+-}$ and $D_{-+}$ can be drawn as the right hand side of Figure~\ref{Dpropagators}. 
All the propagators $D_{++}$, $D_{--}$, $D_{+-}$ and $D_{-+}$ can contribute to the pair production of massive particles. The difference is that, the mixed propagator $D_{+-}$ and $D_{-+}$ can only correspond to the probability of pair production of massive particles, thus their contribution scales as the Boltzmann factor $e^{-2\pi\mu}$. The propagators $D_{++}$ and $D_{--}$ have a leading order contribution which behaves like $\frac{1}{\mu^2}$, corresponding to the vaccum fluctuation of the inflaton field. An alternative viewpoint would be to consider this as a production of a pair of virtual massive $\delta\sigma$ particles. To the next-to-leading order, it has a contribution that behaves like $e^{-\pi\mu}$, corresponding to the amplitude of the pair creation and thus is the square root of the Boltzmann factor. Physically, this could be viewed as an interference effect between the vacuum fluctuation and the pair creation effect.

With this in mind, we are able to estimate the 10 terms which contribute to Figure~\ref{3sigmadiagram} presented in Appendix \ref{concreteform}. We can reformulate them in terms of the Schwinger-Keldysh language and observe what types of massive propagators ($D_{++(--)}$ or $D_{+-(-+)}$) each term contains. Terms (1) and (2) have one propagator of the type $D_{++(--)}$ and two propagators of the type $D_{+-(-+)}$. Term (3) has three propagators of the type $D_{+-(-+)}$. Terms (4), (5) and (6) have two propagators of the type $D_{++(--)}$ and one of the type $D_{+-(-+)}$. Terms (7), (8), (9) and (10) have three propagators of the type $D_{++(--)}$. In this paper, we choose to focus on the large mass behavior which translates to being of the order $e^{-\pi\mu}$, so any term that contains a contribution from the $D_{+-(-+)}$ propagators is subleading because it is at least suppressed by a factor of $e^{-2\pi\mu}$. This means that any diagram that contains at least one massive propagator resembling the type in the left hand side of Figure~\ref{Dpropagators} will be subleading, and only (7), (8), (9) and (10) are left.

\subsection{Partial Effective Field Theory Approach}\label{EoMmethod}
In this subsection, we would like to review the effective field theory approach and apply it to the quasi-single field inflation model, and then extend it to the partial effective field theory approach.

Starting from the Lagrangian of our model \eqref{ModelAction}, we can derive the equation of motion for the massive field as follows,
\begin{equation}\label{eom:deltasigma}
 \ddot {\delta \sigma} +3 H \dot{ \delta \sigma} -\left(\frac{\nabla ^2}{a^2} - m_{\rm eff}^2\right)\delta \sigma +\frac{V'''}{2} {\delta \sigma}^2 = 2 R \dot \theta_0 \dot {\delta \theta}~.
\end{equation}
where the effective mass is defined as $m_{\rm eff} = V''(\sigma_0)-\dot{\theta_0}^2$.
For $\delta\sigma$ is heavy, its derivatives can be neglected compared to the mass term. Therefore \eqref{eom:deltasigma} gives an algebraic relation, which can be solved iteratively:
\begin{align}\label{EOMofthesigma}
 \delta \sigma =\frac{2 R \dot \theta _0}{m_{\rm eff} ^2}\dot {\delta \theta} +\bigg(\frac{R}{{m_{\rm eff} ^2}c_s^2}-\frac{2 R^2 {\dot \theta _0}}{m_{\rm eff} ^2}  \frac{V'''}{m_{\rm eff} ^4}\bigg) {\dot {\delta \theta}}^2+...~.
\end{align}
where the masked terms in ellipsis ($\ldots$) are higher order in $\dot{\delta\theta}$, and $c_s$ is the sound speed defined as
\begin{align}\label{soundspeed}
 \frac{1}{c_s^2} \equiv 1 + \frac{4\dot\theta^2}{m^2_{\rm eff}} ~.
\end{align}
Notice that the algebraic relation \eqref{EOMofthesigma} is local, which means the nonlocal operators like $\partial^{-2}$ has been neglected. This is the reason why the power spectrum and bispectrum derived from EFT approach has only contained the local contributions in the large mass limit.

Now we insert this relation into the $\delta\sigma^3$ interaction. Note, that in order to find the leading-order nonlocal effects, 
we only substitute the algebraic relation \eqref{EOMofthesigma} into two of the three $\delta \sigma $ fields, instead of all of the three $\delta\sigma$'s as in \cite{Gong:2013sma}. This is why our method is called \textit{partial}, since this means only part of the heavy propagators are integrated out. The resulting Lagrangian is
\begin{align}\nonumber
 \mathcal L= a^3\bigg[ & - \frac{R^2}{2 a^2}(\nabla \delta \theta)^2- \frac{1}{2 a^2} (\nabla \delta \sigma)^2 - \frac{R}{a^2} \delta \sigma (\nabla \delta \theta)^2- \frac{1}{2} {m_{\rm eff} ^2} {{\delta \sigma}^2}                                                                                                        \\
                       & + 2R \dot \theta _0 \dot {\delta \theta} \delta \sigma +\frac{2 R {\dot \theta _0}^2}{m_{\rm eff} ^2} {\dot {\delta \theta}}^2 \delta \sigma +R {\dot {\delta \theta}}^2 \delta \sigma -\frac{2}{3} V''' \frac{R^2 {\dot \theta _0}^2}{m_{\rm eff} ^4} {\dot {\delta \theta}}^2 \delta \sigma\bigg]~.
\end{align}
So we have the third order Lagrangian
\begin{align}\label{L3}\nonumber
 \mathcal L_3 & = a^3 \bigg[R \frac{V'' + {\dot \theta_0}^2}{V'' - {\dot \theta_0}^2} -\frac{2}{3} V''' \frac{R^2 {\dot \theta_0}^2}{(V''-{\dot \theta_0}^2)^2}\bigg] {\dot {\delta \theta}}^2 \delta \sigma \\
              & \approx a^3 \bigg(-\frac{2}{3} V''' \frac{R^2 {\dot \theta_0}^2}{V''^2}\bigg){\dot {\delta \theta}}^2 \delta \sigma =C_3 a^3 \dot{\delta\theta}^2 \delta\sigma~.
\end{align}
where in the second step we use the fact that $V'''$\,\,is the only possible origin for large non-Gaussianities, and $V'' \gg {\dot \theta_0}^2$. The second order Lagrangian is
\begin{align}
 \mathcal L_2 = a^3 2R \dot \theta_0 \dot {\delta \theta} \delta \sigma=C_2 a^3 \dot{\delta\theta} \delta\sigma~.
\end{align}
We have defined the coefficient $C_2$ and $C_3$ for convenience,
\begin{align}
 C_2 =2 R \dot \theta_0
 ,\quad
 C_3 =-\frac{2}{3} V''' \frac{R^2 {\dot \theta_0}^2}{V''^2}
\end{align}
Using in-in formalism, we have
\begin{align}\nonumber
 \langle \delta\theta^3 \rangle&
 \supset
 \int_{-\infty}^0 d\tilde \tau_1 \int_{-\infty}^0 d\tau_1
 \langle 0| H_I(\tilde \tau_1) \delta\theta^3 H_I(\tau_1) |0\rangle'
 -2 {\rm Re} \left[ \int_{-\infty}^0 d\tau_1 \int_{-\infty}^{\tau_1} d\tau_2
 \langle 0| \delta\theta^3 H_I(\tau_1) H_I(\tau_2) |0\rangle' \right] \\  \label{NTOintegral1sigma}
                                 & = 2 \times 6 u_{k_3}^* u_{k_1} u_{k_2}|_{\tau=0}
 \left( \int_{-\infty}^0 d\tilde\tau C_2 a^3 v_{k_3} u'_{k_3} \right)
 \left( \int_{-\infty}^0 d\tau C_3 a^2 v_{k_3}^* u'^{*}_{k_1} u'^{*}_{k_2} \right)
 \\ \label{TOintegral1sigma}
                                 & -2\times 6 u_{k_3} u_{k_1} u_{k_2}|_{\tau=0}
 \\ \nonumber
                                 & \times{\rm Re} \left[ \int_{-\infty}^0 d\tau_1 C_2 a^3 v_{k_3} u'^*_{k_3}
 \int_{-\infty}^{\tau_1} d\tau_2 C_3 a^2 v_{k_3}^* u'^{*}_{k_1} u'^{*}_{k_2} + \int_{-\infty}^0 d\tau_1 C_3 a^2 v_{k_3} u'^{*}_{k_1} u'^{*}_{k_2}
 \int_{-\infty}^{\tau_1} d\tau_2 C_2 a^3 v^*_{k_3} u'^*_{k_3} \right]
\end{align}
The non-time ordered integral \eqref{NTOintegral1sigma} and the time ordered integral \eqref{TOintegral1sigma} have already been evaluated in \cite{Chen:2015lza}.  It is shown that the non-time ordered integral is suppressed exponentially as a function of mass, e.g. $\sim e^{-2\pi\mu}$ whereas the time ordered integral is suppressed by power law. We are only interested in the correction up to $e^{-\pi\mu}$, thus our order estimation in Sec.~\ref{order} can be used to pick up the integral worth considering. The non-time ordered integral is too much suppressed 
and hence in the following discussion, we will focus on the time ordered integral \eqref{TOintegral1sigma}, which is
\begin{align}\nonumber
 \eqref{TOintegral1sigma} & =- \frac{ V'''\dot \theta_0^3 H^3}{  k_1k_2 k_3  R^3} \frac{ \pi e^{-\pi \mu}}{12 m_{\rm eff}^4} 
 \times 6
 \\ \nonumber
                          & {\rm Re}\bigg[\int_{-\infty}^0d\tau_1 (-\tau_1)^{-1/2} H^{(1)}_{i\mu}(-k_3\tau_1)e^{i k_3\tau_1}  \int_{-\infty}^{\tau_1} d\tau_2 (-\tau_2)^{3/2} H^{(2)}_{-i\mu} (-k_3\tau_2) e^{i(k_1+k_2)\tau_2} \\ \label{EOM_result}
                          & +\int_{-\infty}^0 d\tau_1 (-\tau_1)^{3/2} H^{(1)}_{i\mu} (-k_3\tau_1) e^{i (k_1+k_2)\tau_1} \int_{-\infty}^{\tau_1} d\tau_2 (-\tau_2)^{-1/2} H^{(2)}_{-i\mu}(-k_3\tau_2) e^{i k_3\tau_2}\bigg]
\end{align}
Notice that the numerical factor of $6$ comes from permutations. An ensuing difference between this expression and that in \cite{Chen:2015lza} not only pertains to the mass dependence from integrating out the massive field, but also that in this paper we don't have permutations. The reason being that when we take the squeezed limit $k_1=k_2\gg k_3$, for each permutation, we need to employ the effective field theory approach for the $k_1$ and $k_2$ leg, while remembering that only the $k_3$ has the long wavelength nonlocal contribution.

\subsection{Large Mass Approximation Method}
In this section, we would like to use a different method to simplify the calculation of the Feynman diagram from the one in the previous section. The idea is that, since we are only interested in the effect up to the order $e^{-\pi\mu}$, we can use the large mass approximation of the Hankel functions in four of the Hankel functions out of the six in each of the ten terms. This is equivalent to integrating out the massive propagators. The large mass approximation we use is
\begin{align}\label{largemass}
 H_{i\mu}^{(1)}(x)\xrightarrow{\rm large\,\, mass} & \sqrt{\frac{2}{\pi \mu}}e^{\pi \mu /2 - i\pi /4} \bigg(\frac{e x}{2 \mu}\bigg)^{i\mu} e^{i \frac{x^2}{4\mu}  }  (1+ \mathcal O(\mu^{-1}))~.
\end{align}
Using this expression, we obtain the following integral
\begin{align}\label{largemassint}\nonumber
   & \int_{-\infty}^{\tau_1} \frac{d\tau}{(-\tau)^{1/2}} H_{-i\mu}^{(2)} (-k \tau) e^{i k\tau}                                                                                                                                \\
   & \xrightarrow{\rm large\,\, mass} - i \sqrt{\frac{2}{\pi\mu^3}} e^{\pi\mu/2} e^{i\pi/4} \left( \frac{2\mu}{e k} \right)^{i\mu} (-\tau_1)^{1/2-i\mu} e^{i(k\tau_1- \frac{k^2 \tau_1^2}{4\mu})} (1+ \mathcal O(\mu^{-1}))~.
\end{align}

As we already estimated in Section \ref{order}, we can figure out that up to $\frac{1}{\mu^4}e^{-\pi\mu}$, only the terms (7), (8), (9) and (10) contributes, which gives 
\begin{align}\nonumber
        & (7)+ (8)+(9)+(10) =-\frac{e^{-\pi \mu}\pi }{12 H R^3}\frac{ \dot \theta_0^3 V'''}{k_1 k_2 k_3} \frac{1}{\mu^4}
 \times 6
 \\ \nonumber
 \times & {\rm Re}\bigg[\int_{-\infty}^{0} d \tau_1 (- \tau_1)^{-1/2} H_{i\mu}^{(1)} (-k_3 \tau_1)e^{i k_3 \tau_1} \int_{-\infty}^{\tau_1} d \tau_2  { (-\tau_2 )^{3/2} H_{-i\mu}^{(2)} (-k_3 \tau_2)  e^{i \tau_2  (k_1+k_2)}}   \\ \label{LM_result}
        & +\int_{-\infty}^{0} d \tau_1  (-\tau_1 )^{3/2}  H_{i\mu}^{(1)} (-k_3 \tau_1)   { e^{i  (k_1+k_2) \tau_1 }}  \int_{-\infty}^{\tau_1} d \tau_2 (-\tau_2)^{-1/2}  H_{-i\mu}^{(2)} (-k_3 \tau_2) e^{i k_3\tau_2} \bigg]  ~.
\end{align}
We strip the momentum conservation $ (2\pi)^3  \delta^{(3)}( {\boldsymbol k}_1+ {\boldsymbol k}_2+{\boldsymbol k}_3)$ for clearance, while the reader should be kept in mind that all the three-point functions will be multiplied by a $(2\pi)^3\delta^{(3)}(\mathbf{k}_1+\mathbf{k}_2+\mathbf{k}_3)$ factor.

When $\mu$ is large, $m_{\rm eff}\rightarrow \mu H,~$
the partial EFT result \eqref{EOM_result} agrees with the result obtained by the large mass approximation method. Once we get this expression, it is convenient to use the method introduced in \cite{Chen:2015lza}. Again we have the difference compared with \cite{Chen:2015lza} due to the fact that we are always using the large mass approximation for $\delta\sigma$ propagator on $k_1$ and $k_2$ legs.

\begin{figure}[htbp]
 \centering
 \includegraphics[width=0.47\textwidth]{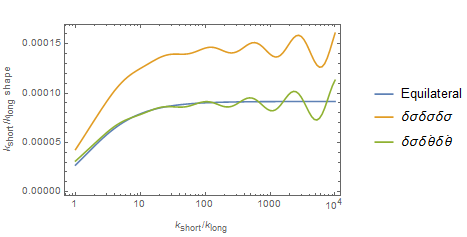}
 \includegraphics[width=0.47\textwidth]{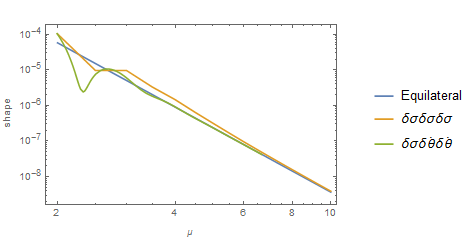}
 \caption{\label{fig:shape} The left panel plots the comparison between the diagram with the effective $\delta\sigma\dot{\delta\theta}\dot{\delta\theta}$ interaction, the diagram with $\delta\sigma^3$ interaction and the equilateral shape obtained in \cite{Gong:2013sma}. It is a plot in momentum space with the ratio $k_1/k_3$ being varied while $\mu=4$. The y axis is the shape function $S(k_1, k_2, k_3)$ magnified by $k_1/k_3$. The right panel is the shape function as a function of the mass parameter $\mu$. In both figures, we choose the parameters $V''' R/H^2 =6 $ and $\dot{\theta_0}/H = 1/2$. }
 \label{3sigma1sigma}
\end{figure}

Next we would like to make a numerical check of whether our new partial effective field theory works. We compare the numerical result obtained by directly evaluating the diagram with $\delta\sigma^3$ interaction and the numerical result of the partial effective field theory diagram with the $\delta\sigma\dot{\delta\theta}\dot{\delta\theta}$ interaction. If one integrates out all the $\delta\sigma$ propagators, one will get an equilateral shape non Gaussianity \cite{Gong:2013sma}.
\begin{align}
 \langle \delta \theta^3 \rangle_{\rm eq} = - \frac{4\dot{\theta_0}^3}{\mu^6 H R^3} \frac{V'''}{k_1 k_2 k_3 (k_1 + k_2 + k_3)^3}~.
\end{align}
It is convenient to define a dimensionless quantity $S(k_1,k_2,k_3)$
\begin{align}
 S(k_1,k_2,k_3) = (k_1 k_2 k_3)^2 \langle \delta \theta^3 \rangle~.
\end{align}
The results are presented in Figure~\ref{3sigma1sigma}. From the plot, we can see that our partial effective field theory approach is useful to obtain the leading-order nonlocal effects, i.e. the ``standard clock'' signal in the squeezed limit $k_1 = k_2 \gg k_3$. The displacement from the $\delta\sigma\delta\sigma\delta\sigma$ numerical result and the partial EFT result is from the next-to-leading order $(1/\mu)$-corrections in \eqref{largemass}, which will decrease significantly when $\mu$ becomes large. 
However the nonlocal signal which serves as the clock signal in the bispectrum also shrinks when $\mu$ increases, which leaves all three curves merge in the large mass limit. To show the leading-order nonlocal signal, we should pick a  $\mu$ in an intermediate range.
Also, in the non-squeezed configuration, we obtain the leading terms of both local and nonlocal signals. However, due to the fact that in the non-squeezed limit the local signal is much larger, the nonlocal signal gets submerged. Only when we are in the very squeezed limit does the local signal becomes damped because of the equilateral shape configuration, and the nonlocal clock signal emerges manifestly only in this limit.

\section{Strongly Coupled System}\label{Strongly}
In this section we try to extend our result in the previous section to the strong coupled quasi-single field inflation when the coupling $\dot{\theta_0}$ is large. To do so, we introduce three classes of methods to approach this scenario.
\begin{itemize}
 \item Numerically solving the equation of motion. This method can take into account both the local and non-local contributions and it is equivalent to considering all the Schwinger-Keldysh propagators up to all orders in \cite{Chen:2015dga}. However it is worth nothing that this is solely a numerical result and cannot be employed to calculate the bispectrum.
 \item If we neglect the non-local contributions, we are able to use the effective field theory method to analytically evaluate the power spectrum and bispectrum. This method is easy for calculations, but the limitation is that it is only accurate in the large mass limit, and nonlocal signals are neglected. We can extend it by preserve one heavy propagator unintegrated out which we dubbed as the partial EFT, to include the leading-order nonlocal signal.
 \item We can verify EFT by using the Schwinger-Keldysh diagrammatics developed recently in \cite{Chen:2017ryl}. It is confirmed that if we neglect the non-local contributions, the large mass limit of the SK method is equivalent to the effective field theory method neglecting the non-local contributions.
\end{itemize}

\subsection{Equation of Motion Method for Numerical Study}
Now we outline how to solve the equations of motion for numerical study, as to check our final result by SK diagrammatics. We consider a two-field inflationary model and denote the two fields as $\phi_a(\boldsymbol x, t), a = 1, 2$. So the quadratic part of the Lagrangian can be written as
\begin{align}
 \tilde L (\delta \phi_a, \delta \dot{\phi}_a, t) \equiv L(\phi_a, \dot{\phi}_a) - L (\bar\phi_a, \dot{\bar{\phi}}_a) - \int d^3 x \frac{\partial L}{\partial \bar{\phi}_a } \delta \phi_a - \int d^3 x \frac{\partial L}{\partial \dot{\bar{\phi}}_a} \delta {\dot{\phi}_a}~.
\end{align}
The equation of motion is
\begin{align}
 \frac{d}{dt} \left( \frac{\partial \tilde L}{\partial \delta \dot{\phi}_a } \right) - \frac{\partial \tilde L}{\partial \delta \phi_a} = 0, \quad a = 1,2~.
\end{align}
We consider the initial time $t_0$,
\begin{align}
 u_a (\boldsymbol k, t_0) = u^{\rm ini}_a (\boldsymbol k)~, \quad \dot{u_a} (\boldsymbol k, t_0) = \tilde u^{\rm ini}_a (\boldsymbol k)~,
\end{align}
while the initial condition satisfies the canonical kinetic relation
\begin{align}
 a^3 (t_0) (u_a \tilde u_a^* - {\rm c.c.} ) = i , \quad (\text{no sum over $a$})~.
\end{align}
Then we need to solve the equation of motion twice with different sets of initial conditions
\begin{align}
 u_1^{(1)} = u_1^{\rm ini}, \quad \dot{u}_1^{(1)} = \tilde u_1^{\rm ini}; \quad u_2^{(1)} = 0,\quad u_2^{(1)} = 0~.
\end{align}
and
\begin{align}
 u_2^{(2)} = u_2^{\rm ini}, \quad \dot{u}_2^{(2)} = \tilde u_2^{\rm ini}; \quad u_1^{(2)} = 0,\quad u_1^{(2)} = 0~.
\end{align}
We get two different sets of solutions $u_a^{(\alpha)}(\boldsymbol k, t)$, where $\alpha, a = 1, 2$.
\begin{align}
 \langle \delta \phi_{a\boldsymbol p} (t) \delta \phi_{b\boldsymbol q} (t) \rangle = (2\pi)^3 \sum_{\alpha=1}^2 u_a^{(\alpha)} (\boldsymbol p, t) u_b^{(\alpha) *} (\boldsymbol q, t) \delta^3 (\boldsymbol p+ \boldsymbol q)~.
\end{align}
If we apply this formalism to a system with one light field and one heavy field (quasi-single field inflation, for instance), and the mass of the heavy field is large enough to make the non-local contributions negligible, we can then integrate out the heavy degree of freedom, and get the effective mode function of the light one analytically which corresponds to the usual Bunch-Davies mode function with a modification from the sound speed
\begin{align}
 u_{k}^{\rm{(eff)}} (\tau) = \frac{H}{R \sqrt{2 c_s k^3}} (1+ i c_s k \tau) e^{- i c_s k \tau}
\end{align}
where the expression for the sound speed is in \eqref{soundspeed}.

\subsection{Schwinger-Keldysh Diagram}
In this section, we use the Schwinger-Keldysh diagram to calculate the effective mode function.  We need to calculate the effective bulk-to-boundary propagator $\langle \delta\phi(\tau)\delta\phi(0)\rangle$. To do this, we calculate the first order propagator and insert it into the second order one and so on. Here we present how to obtain the $(n+1)$-th order propagator provided that the form of the $n$-th order is known.

\begin{figure}[htbp]
 \centering
 \includegraphics[width=1\textwidth]{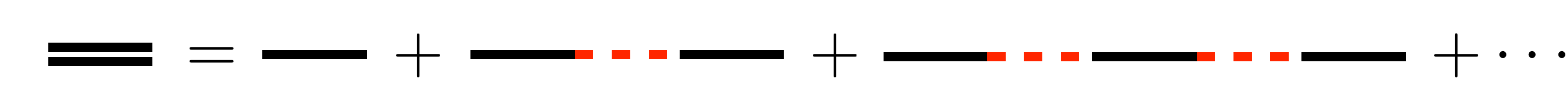}
 \caption{\label{fig:shape}  This is a diagramatic way to show how to use the SK diagram method to calculate the effective mode function.}
 \label{propagators}
\end{figure}

We first write down the formulae for the general propagators. We denote $\mathcal G_{\pm(n)} (k,\tau)$ as the propagator with one heavy leg and one light leg, while $\mathcal U_{\pm (n)} (k;\tau)$ is the propagator with two light legs, with subscripts the numbers of their internal massive propagators. Therefore, by intuitive deduction, we have
\begin{align}
 \mathcal G_{\pm(n+1)} (k,\tau)  & = i 2 \dot{\theta_0} \int_{-\infty}^0 \frac{d\tau'}{(-H\tau')^3} \bigg[ D_{\pm + } (k;\tau,\tau') \partial_{\tau'} \mathcal U_{+(n)} (k;\tau') -D_{+-} (k;\tau,\tau') \partial_{\tau'} \mathcal U_{-(n)} (k,\tau')  \bigg], \\
 \mathcal U_{\pm (n+1)} (k;\tau) & = i 2 \dot{\theta_0} \int_{-\infty}^0 \frac{d\tau'}{(-H\tau')^3} [\partial_{\tau'}G_{\pm+}(\tau,\tau')\mathcal G_{+(n+1)}(k,\tau')-\partial_{\tau'}G_{\pm-}(\tau,\tau')\mathcal G_{-(n+1)}(k,\tau') ] .
\end{align}
By using the above formulae we can sum up all propagators with two light legs, and then push one of the final leg to the boundary, which can be serve as an effective boundary-to-bulk propagator as is shown in Figure~\ref{propagators},
\begin{align}
\mathcal U_+^{\text{(eff)}} (z) &= \sum_{n=0}^\infty \mathcal U_{+(n)} (z) \\\nonumber
                              & = \frac{H^2 e^{i z} (1-i z)}{2 k^3 R^2}+\frac{(2 \dot{\theta_0})^2 e^{i z} (1-z (z+i))}{4 k^3 \mu ^2 R^4}+\frac{(2 \dot{\theta_0})^4 e^{i z} (-1+i z (1+z (z-2 i)))}{16 H^2 k^3 \mu ^4
 R^6}\\ \nonumber
                              & +\frac{(2 \dot{\theta_0})^6 e^{i z} (3+z (z (-9+z (z-8 i))-3 i))}{96 H^4 k^3 \mu ^6 R^8}+\frac{(2 \dot{\theta_0})^8 e^{i z} (-15+z (z (60-i z (-73+z (z-17 i)))+15 i))}{768 H^6 k^3
 \mu ^8 R^{10}}\\
                              & +\frac{(2 \dot{\theta_0})^{10} e^{i z} (105-z (z (525+z (z (-260+z (z-29 i))+790 i))+105 i))}{7680 H^8 k^3 \mu ^{10} R^{12}}+O\left(\left(\frac{\dot\theta_0}{H\mu}\right)^{12}\right)\label{summation}
\end{align}

We can see from this expression that the $n$-th order term is of order $\dot{\theta_0}^{2n}/\mu^{2n}$. This is because each transfer vertex contributes a factor of $\dot{\theta_0}$, and each propagator, in the large mass limit, contributes $\frac{1}{\mu^2}$. The error obtained by this method is $\mathcal O (\frac{1}{\mu})$. 
We can see that as $\dot{\theta_0}$ increases, the error decreases as expected. 
The power spectrum $\langle \delta\theta\delta \theta\rangle$ can be obtained by taking the summation of \eqref{summation} after pushing another leg to the boundary, i.e. taking the $z\rightarrow0$ limit:
\begin{align}
 \langle \delta\theta (0) \delta\theta(0) \rangle = u_{\rm eff} (0) u^*_{\rm eff} (0) =\mathcal U_+^{\text{(eff)}} (0) =  \frac{H^2}{2R^2 k^3} \sqrt{1+\frac{4\dot\theta_0^2}{H^2\mu^2}}= \frac{H^2}{2R^2c_s k^3}
\end{align}
It is consistent with the first order result in \cite{Chen:2012ge,Pi:2012gf}, and the full-order result in \cite{Cremonini:2010ua}, which means that we can safely extrapolate the first order result to all orders, as well as the coupling $\dot\theta_0$ is not \textit{too} large as we will discuss below. A remark of this is that when using the effective field theory carefully \cite{Tong:2017iat}, the result is
\begin{align}
 \langle \delta\theta (0) \delta\theta(0) \rangle = u_{\rm eff} (0) u^*_{\rm eff} (0) = \frac{H^2}{2R^2c_s k^3} =  \frac{H^2}{2R^2 k^3} \sqrt{1+\frac{4\dot\theta_0^2}{m^2-2H^2}}
\end{align}
Since we are focusing on the large mass parameter regime, the difference of the constant is not important.

\begin{figure}[htbp]
 \centering
 \includegraphics[width=0.7\textwidth]{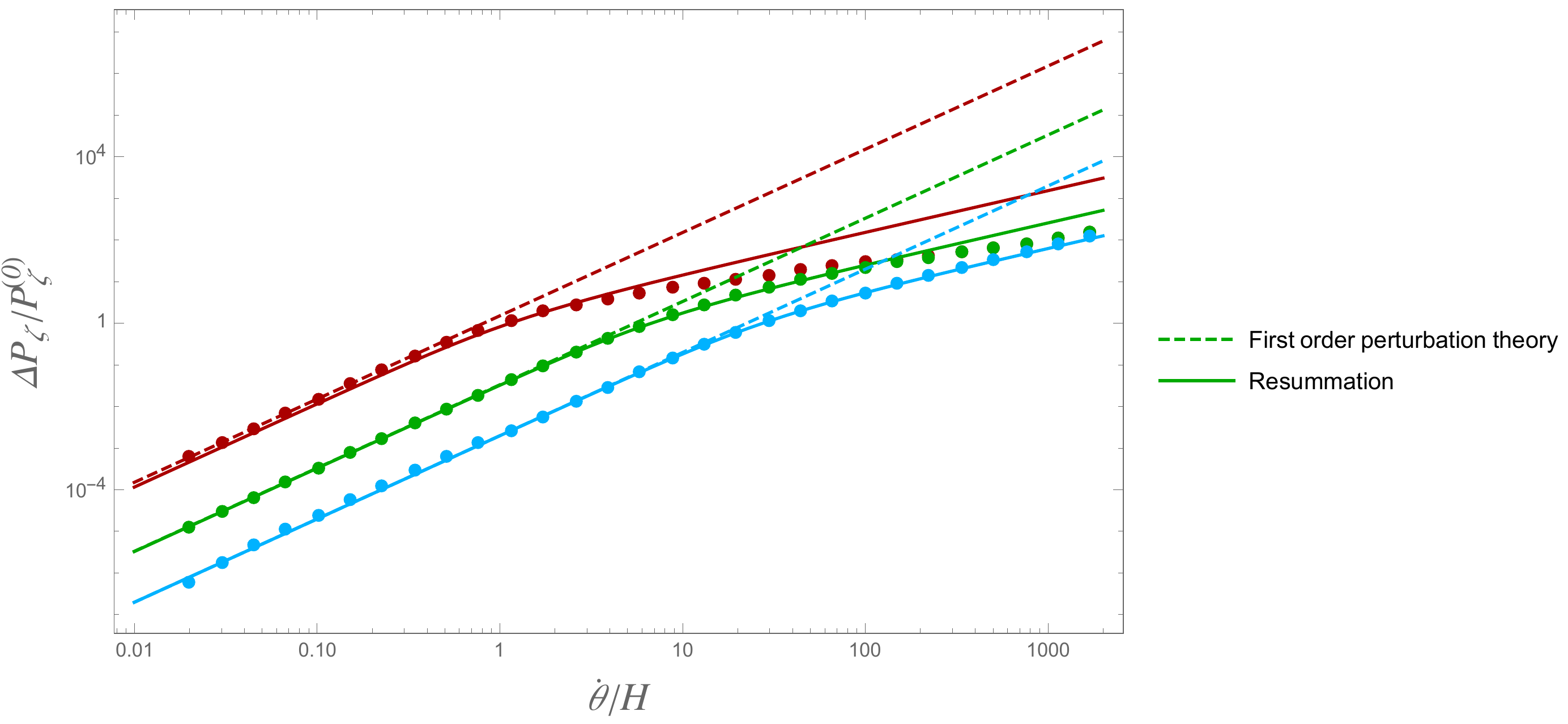}
 \caption[aaaa]{This plot shows the range of applicability of the first order perturbation theory and the resummation technique. The dotted lines are the numerical solutions obtained from directly solving the coupled differential equations. The dashed lines are from the first order perturbation theory. The solid lines are results obtained by our resummation technique. Red, Green and Cyan represents $m_{\rm eff} = 2, 8, 32$, respectively.  \protect\footnotemark} \label{aaaa}
\end{figure}

In Figure~\ref{aaaa}, we plot the relative power spectrum deviation $\Delta P_\zeta/P_\zeta^{(0)}$ of the quasi-single field inflation model from the original single field inflation model as a function of the coupling between the inflaton and the massive field. The power spectrum is defined in terms of the two point correlation function in the standard way as
\begin{align}
P_\zeta = \frac{H^2}{R^2\dot\theta_0^2} \frac{k^3}{2\pi^2} \langle \theta(0) \theta(0) \rangle
\end{align}
$P_\zeta^{(0)}$ is the power spectrum of the original single field inflation model without massive field. 
We can see that the first order perturbation theory is applicable in the parameter range $\dot \theta_0 <m_{\rm eff}$ and the resummation technique is applicable in the parameter range $\dot \theta_0 H<m^2_{\rm eff}$.

\footnotetext{We thank Xi Tong for drawing this plot during the discussion of the work \cite{Tong:2017iat}.}

\section{Sound speed corrected bispectrum}\label{correctedbispectrum}
\begin{figure}[htbp]
 \centering
 \includegraphics[width=0.4\textwidth]{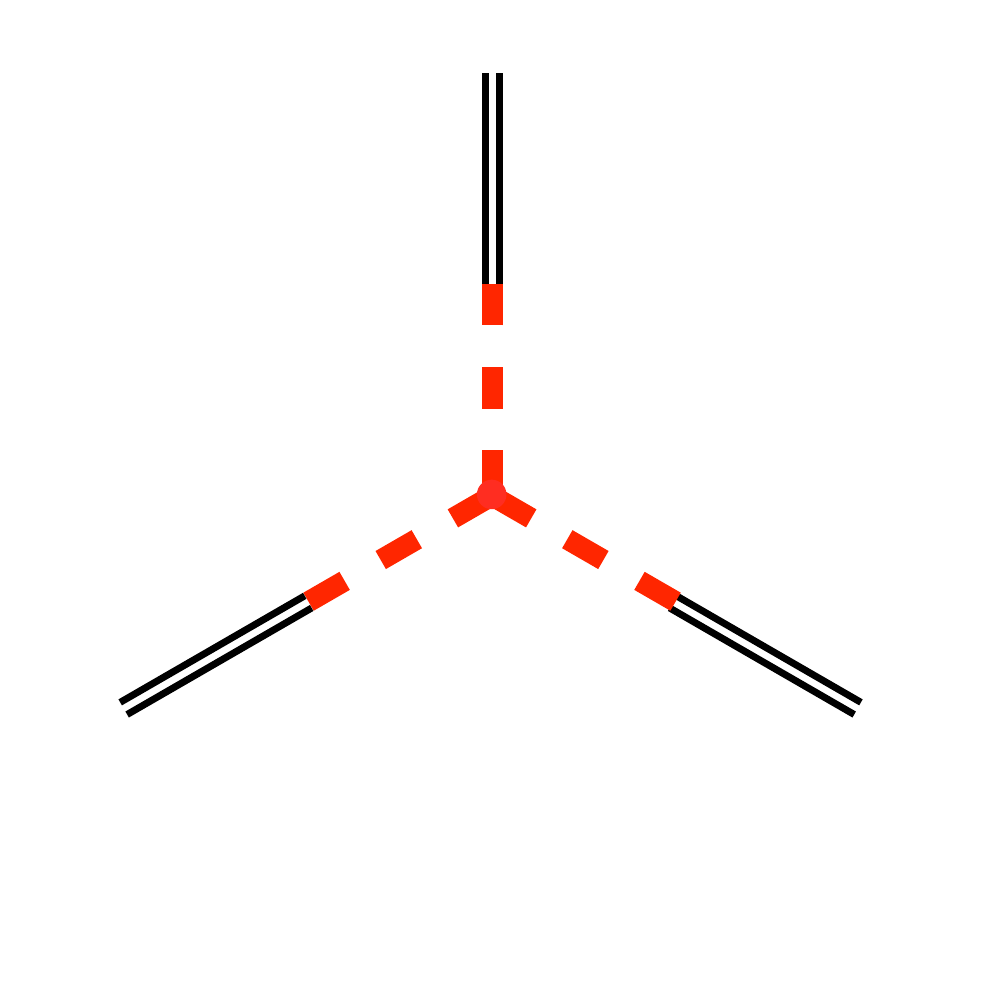}
 \includegraphics[width=0.4\textwidth]{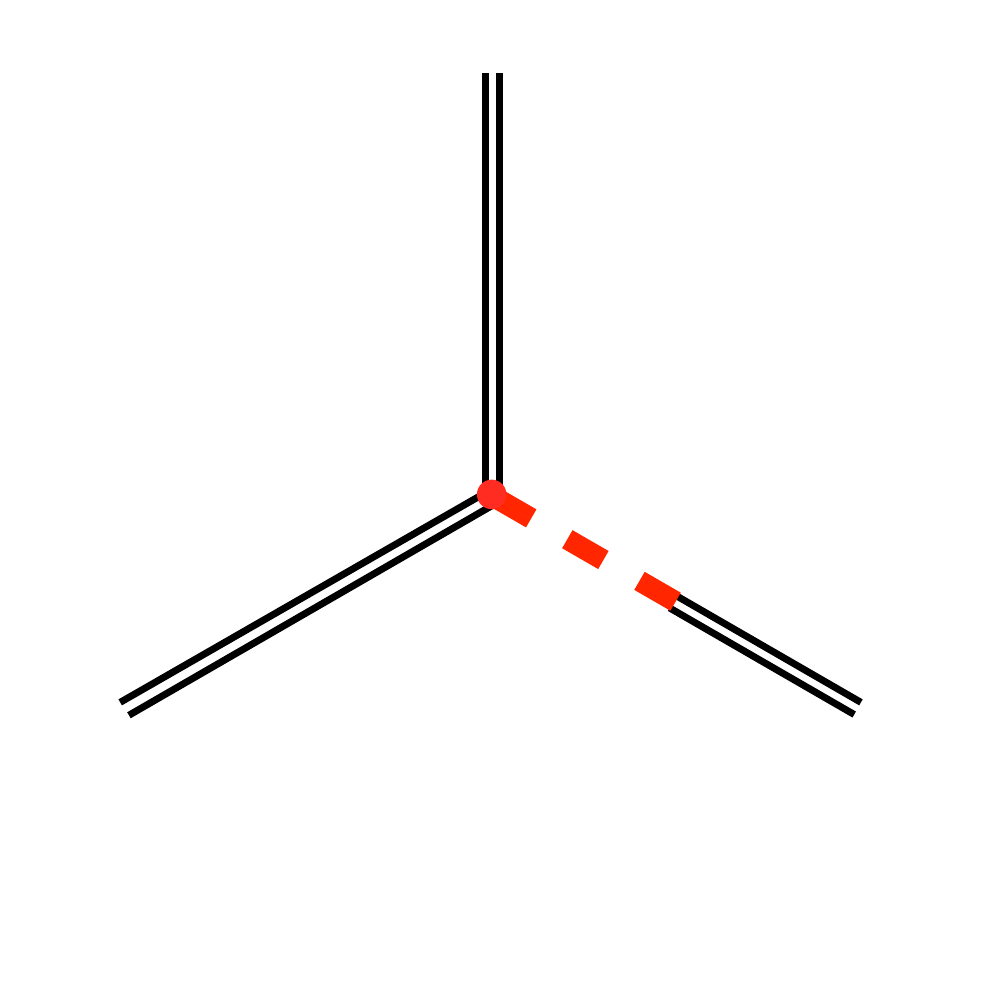}
 \caption{ These are two types of Feynman diagram we get from the strongly coupled quasi single field inflation. The Feynman diagram on the left hand side corresponds to the $\delta\sigma^3$ interaction and the diagram on the right hand side corresponds to the $\dot{\delta\theta}\dot{\delta\theta}\delta\sigma$ interaction.  }
\label{fig:strong} 
\end{figure}

Equipped with the effective mode function presented in the previous sections, we can start to calculate the bispectrum in the weekly coupled regime with a modification of the sound speed. We consider the diagram with $\delta\sigma^3$ interaction shown on the left hand side of Figure~\ref{fig:strong}. It can be obtained by first using the partial effective field theory approach and then using the effective mode function. The diagram with $\dot{\delta\theta}\dot{\delta\theta}\delta\sigma$ interaction can be obtained analogously by only replacing the massless mode function into the effective mode function as shown on the right hand side of Figure~\ref{fig:strong}. Inserting the effective mode functions in \eqref{NTOintegral1sigma} and \eqref{TOintegral1sigma}, we obtain
\begin{align} \nonumber
 \langle \delta\theta^3 \rangle & = \frac{\pi  c_s^3 H^3 e^{-\pi  \mu } C_2 C_3 }{16 k_1 k_2 k_3^4 R^6} \bigg[ - \int_{0}^{+\infty} d\tilde z  {\tilde z}^{-1/2} H^{(1)}_{i\mu} (\tilde z)  e^{i c_s \tilde z} \int_{0}^{+\infty} d z  z^{3/2} H^{(2)}_{-i\mu} (z) e^{-i c_s \frac{ k_1+k_2}{k_3} z} \\   \nonumber
                                & + \int_{0}^{+\infty} dz_1 z_1^{-1/2} H^{(1)}_{i\mu} (z_1) e^{-i c_s z_1} \int_{z_1}^{+\infty} dz_2 z_2^{3/2} H^{(2)}_{-i\mu} (z_2) e^{-i c_s \frac{k_1+k_2}{k_3} z_2}                                                                                              \\ \label{51}
                                & + \int_{0}^{+\infty} d z_1 z_1^{3/2} H^{(1)}_{i\mu}(z_1) e^{ - i c_s  \frac{k_1 + k_2}{k_3} z_1} \int_{z_1}^{+\infty} dz_2 z_2^{-1/2} H^{(2)}_{-i\mu} (z_2) e^{-i c_s z_2} \bigg]+ {\rm c.c.}+ {\rm 2~perm.} ~,
\end{align}
Firstly, we calculate the first layer of integral. Consider the squeezed limit $k_3\ll c_s (k_1+k_2)$, we can expand the Hankel functions at IR. We affix the details of the calculation of the integral in Appendix \ref{integrals}. The incomplete Gamma function terms will be suppressed, resulting in the final result of the three-point function, including the prefactor computed as follows,

\begin{align}\nonumber
 \langle \delta\theta^3 \rangle= \frac{\pi  c_s^3 H^3 e^{-\pi  \mu } C_2 C_3 }{16 k_1 k_2 k_3 R^6} \bigg[ & \Gamma\bigg( \frac{5}{2} -i\mu \bigg)   \bigg( \frac{i c_s (k_1+k_2)}{k_3} \bigg)^{-\frac{5}{2}+i\mu} \left(  \frac{2^{i \mu }  (\coth (\pi  \mu )+1) }{\Gamma (1-i \mu )} \mathcal I - \frac{i 2^{i \mu } \Gamma (i \mu )  }{\pi } \mathcal I^*  \right) \\ \nonumber
 +                                                                                                        & \Gamma\bigg( \frac{5}{2} +i\mu \bigg)    \bigg( \frac{i c_s (k_1+k_2)}{k_3} \bigg)^{-\frac{5}{2}-i\mu} \left( \frac{i 2^{-i \mu } \Gamma (-i \mu ) }{\pi } \mathcal I +   \frac{2^{-i \mu } (\coth (\pi  \mu )+1) }{\Gamma (i \mu
 +1)} \mathcal I^* \right) \bigg] \\\label{finalresult}
 +                                                                                                        & {\rm c.c.} + {\rm 2\,\,\, Perms}
\end{align}
where the $\mathcal I$ is defined as,
\begin{align}\nonumber
 \mathcal I & \equiv \int_{0}^{+\infty} d\tilde z  {\tilde z}^{-1/2} H^{(1)}_{i\mu} (\tilde z)  e^{i c_s \tilde z}                                                          \\
            & = e^{\pi\mu/2} \frac{(i/2)^{-1/2}}{\sqrt{\pi}} \Gamma(\frac{1}{2}-i\mu)\Gamma(\frac{1}{2}+i\mu) {}_2F_1 (\frac{1}{2}-i\mu,\frac{1}{2}+i\mu,1,\frac{1-c_s}{2})
\end{align}
We do an order estimate here for large $\mu$,
\begin{align}
   & \Gamma(-i\mu) \Gamma(i\mu+\frac{5}{2} ) \sim e^{-\pi\mu}, \quad  \frac{2^{i \mu }  (\coth (\pi  \mu )+1) }{\Gamma (1-i \mu )}  \Gamma \left(\frac{5}{2}-i \mu \right)\sim 1, \quad \mathcal I\sim \mathcal I^* \sim e^{-\frac{\pi\mu}{2}} \\
   & \frac{2^{-i \mu } (\coth (\pi  \mu )+1) }{\Gamma (i \mu
 +1)}   \Gamma\bigg( \frac{5}{2} +i\mu \bigg)  \sim 1 ,\quad  -\frac{i 2^{i \mu } \Gamma (i \mu )  }{\pi }   \Gamma\bigg( \frac{5}{2} -i\mu \bigg) \sim e^{-\pi\mu}
\end{align}
Put everything together, we get
\begin{align}\nonumber
 {\rm non\,\,\, time \,\,\,ordered\,\,\,contribution} & = \# e^{-2\pi\mu} \left(\frac{ c_s
 \left(k_1+k_2\right)}{k_3}\right)^{-\frac{5}{2}+i \mu } + \# e^{-2\pi\mu}  \left(\frac{ c_s
 \left(k_1+k_2\right)}{k_3}\right)^{-\frac{5}{2}-i \mu }
\end{align}
\begin{align}\nonumber
 {\rm time \,\,\,ordered\,\,\,contribution} & =  \# e^{-\pi\mu} \left(\frac{ c_s
 \left(k_1+k_2\right)}{k_3}\right)^{-\frac{5}{2}-i \mu } + \# e^{-3\pi\mu} \bigg( \frac{ c_s (k_1+k_2)}{k_3} \bigg)^{-\frac{5}{2}+i\mu} \\
                                            & \sim  \# e^{-\pi\mu} \left(\frac{ c_s
 \left(k_1+k_2\right)}{k_3}\right)^{-\frac{5}{2}-i \mu }
\end{align}
where the $\#$ denotes some prefactors of power-law or polynomial in $\mu$. 
If we take into account the complex conjugate part in \eqref{51}, the final result, up to the leading order in $\mu$, can be written as
\begin{align}
 \langle \delta\theta^3\rangle\sim \# e^{-\pi\mu} \left(\frac{ c_s
 \left(k_1+k_2\right)}{k_3}\right)^{-\frac{5}{2}-i \mu }\frac1{k_1k_2k_3} +\# e^{-\pi\mu} \left(\frac{ c_s
 \left(k_1+k_2\right)}{k_3}\right)^{-\frac{5}{2}+i \mu } \frac1{k_1k_2k_3}.
\end{align}

\begin{figure}[htbp]
 \centering
 \includegraphics[width=0.4\textwidth]{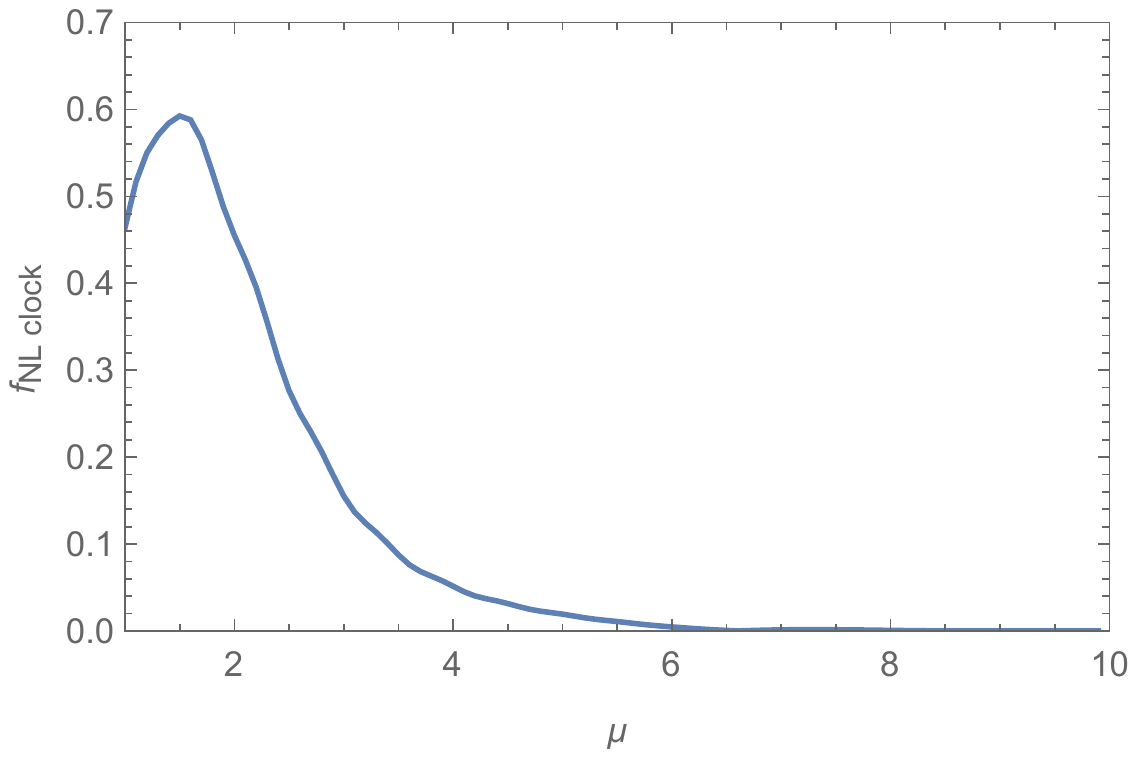}
 \caption{This figure shows the magnitude of the clock signal as a function of the mass parameter $\mu$ when the equilateral non-Gaussianity is fixed to be $1$. We take $P_{\zeta} = 2.2\times 10^{-9}$, $V'''R/H^2 = 1$.   }
 \label{fnl}
\end{figure}

To show the amplitude of the non-Gaussianity from \eqref{finalresult}, we invoke the equilateral non-Gaussianity in the large mass limit~\cite{Gong:2013sma}
\begin{align}
 \langle \zeta \zeta \zeta \rangle & = (2\pi)^7 \delta^{(3)} (\boldsymbol k_1 + \boldsymbol k_2 + \boldsymbol k_3) P_{\zeta}^2 \frac{S_{\delta\sigma^3}(k_1,k_2,k_3)}{k_1^2 k_2^2 k_3^2} \\
 S_{\delta\sigma^3} (k_1,k_2,k_3)  & = \frac{4\pi^6 R \dot{\theta_0}^4 c_s^2}{\mu^6 H^6} V''' \frac{k_1 k_2 k_3}{(k_1+k_2+k_3)^3}                                                        \\
 f_{\rm NL eq}^{\delta\sigma^3}    & = \frac{40}{243} \frac{R \dot{\theta_0}^4c_s^2}{\mu^6H^6} V'''
\end{align}
and use the results as already known (best-fit Planck result for instance) to choose the parameters in \eqref{finalresult}. We draw the nonlinear parameter $f_\text{NL}$ from \eqref{finalresult} under some typical parameter choice in Figure~\ref{fnl}.

A side remark is that using the method introduced in \cite{An:2017hlx}, we can get the exact scaling behavior of the bispectrum $\sim (k_{1}/k_3)^{i \sqrt{m^2/H^2+\dot{\theta}_0^2/H^2}}$. This scaling is true for all the parameter spaces of $m^2/H^2+\dot{\theta}_0^2/H^2>9/4$. But the prefactor should be obtained via fitting the numerical solution with the scaling behavior. Our method does not get exact scaling behavior. Our result is applicable in the regime $\dot \theta_0/H<m/H$, this scaling is the same as that of \cite{An:2017hlx}. In this regime, the sound speed $c_s$ does not deviate too much from 1.

\section{Conclusion and Outlook}
\setcounter{equation}{0}
We have simplified the calculations of the $\delta\sigma^3$-induced three-point function of the curvature perturbation in the quasi-single field inflation significantly by reducing it to $\delta\sigma\dot{\delta\theta}\dot{\delta\theta}$ interaction using a partial effective field theory approach. One can separate the contribution of the massive field into a vacuum part and a thermal part. Here, the vacuum contribution comes from the part without particle production while the thermal contribution counts for the production of one, two or three pairs of heavy particles and so on. For the short wavelength mode, the dominant contribution is of the vacuum type, which corresponds to a virtual particle production scenario. For this particular kind of massive field, we can utilize the large mass approximation or the effective field theory to integrate them out. For the long wavelength mode, we retain it to reproduce the clock signal in the squeezed limit.

The (partial) effective field theory approach can also be extended from weak coupling case with $\dot\theta_0^2/H^2\ll1$ to the ``strong coupling'' regime with $\dot \theta_0/H>1$ while still keeping $\dot \theta_0<m_{\rm eff}^2/H$. Using the resummation technique, we can explicitly show that this amounts to the single field inflation with a modification of sound speed $c_s$. The non-trivial oscillating behavior in the squeezed limit can also be obtained by using a combination of the partial effective field theory technique and the resummation technique.

There are several interesting directions to look at in the future. First, a more comprehensive study of the strongly coupled bispectrum is needed. This includes a more careful method to obtain the exact scaling behavior of the squeezed limit and its prefactor characterizing the amplitude of nonlocal effects.

The partial effective field theory approach may also be applied to the loop diagram computation. For example, if one is only interested in the leading nontrivial cosmological collider type of signal produced by a loop diagram with one loop consisted of three heavy propagators, one can treat two of the interaction vertices connected to a heavy mode with short wavelength as one effective interaction vertex by effective field theory. We can also use the large mass approximation of this propagator that is in between the two short-wavelength external legs. In this case, the diagram simplifies enabling easier extraction of the clock signals.

\section*{Acknowledgments}
We thank Xingang Chen, Gong Cheng, Asuka Ito, Yubin Li, Toshifumi Noumi, Misao Sasaki, Jiro Soda, Chon-Man Sou, Xi Tong, Dong-Gang Wang, and Jun'ich Yokoyama for interesting discussions. We thank Ziyan Yang for her initial collaboration at the early stage of this work. YW is supported by grants HKUST4/CRF/13G, GRF Grant 16301917 and ECS 26300316 issued by the Research Grants Council of Hong Kong. SZ is supported by the Hong Kong PhD Fellowship Scheme (HKPFS) issued by the Research Grants Council (RGC) of Hong Kong. AVI is supported by the Targeted Scholarship Scheme under the HKSAR Government Scholarship Fund. SP is supported by MEXT KAKENHI No. 15H05888. We would also like to thank the Institute for Advanced Study, Hong Kong University of Science and Technology, where part of this work is finished. SZ and AVI also express their gratitude to McGill University for providing a conducive atmosphere vital in shaping key elements of this work. ZW would like to thank the Institute for Advanced Study, Hong Kong University of Science and Technology for the hospitality where he did his undergraduate thesis.

\appendix
\section{Concrete Form of the 10 Terms}\label{concreteform}
In this section, we present the concrete form of the 10 terms in \eqref{FacForm_term}. The 10 terms are denoted (1), (2), \ldots ,(10), respectively.
\begin{align}\nonumber
 (1)= & -12 c^3_2 c_3 u^*_{k_1} (0) u_{k_2}(0)u_{k_3}(0)\times {\rm Re} \bigg[\frac{e^{-3\pi \mu}\pi ^{3/2} \sqrt{k_1 k_2 k_3}}{16 \sqrt{2} H^4 R^3}                                                                                                                                                                                     \\\nonumber
      & \times \int_{-\infty}^0 d \tilde{\tau_1} \frac{1}{\sqrt{- \tilde{\tau_1}}} H_{-i\mu}^{(2)} (-k_1 \tilde{\tau_1}) e^{-i k_1 \tilde{\tau_1}} \int_{-\infty}^{\tilde{\tau_1}} d \tilde{\tau_2} \sqrt{- \tilde{\tau_2}} H_{i\mu}^{(1)} (-k_1 \tilde{\tau_2})H_{i\mu}^{(1)} (-k_2 \tilde{\tau_2})H_{i\mu}^{(1)} (-k_3 \tilde{\tau_2}) \\\nonumber
      & \times \int_{-\infty}^0 d \tau_1 \frac{1}{\sqrt{- \tau_1}} H_{-i\mu}^{(2)} (-k_2 \tau_1) e^{i k_2 \tau_1} \int_{-\infty}^{\tau_1} d \tau_2 \frac{1}{\sqrt{- \tau_2}} H_{-i\mu}^{(2)} (-k_3 \tau_2) e^{i k_3 \tau_2} \bigg]                                                                                                       \\
      & \times  (2\pi)^3  \delta^3  ( {\boldsymbol k}_1+ {\boldsymbol k}_2+{\boldsymbol k}_3)+5 \,\, {\rm perm.}
\end{align}
\begin{align}\nonumber
 (2)= & -12 c^3_2 c_3 u^*_{k_1} (0) u_{k_2}(0)u_{k_3}(0)\times {\rm Re} \bigg[\frac{e^{-3\pi \mu}\pi ^{3/2} \sqrt{k_1 k_2 k_3}}{16 \sqrt{2} H^4 R^3}                                                                                                                                                                                         \\\nonumber
      & \times \int_{-\infty}^0 d \tilde{\tau_1} \sqrt{- \tilde{\tau_1}} H_{-i\mu}^{(2)} (-k_1 \tilde{\tau_1}) H_{i\mu}^{(1)} (-k_2 \tilde{\tau_1}) H_{i\mu}^{(1)} (-k_3 \tilde{\tau_1})  \int_{-\infty}^{\tilde{\tau_1}} d \tilde{\tau_2} \frac{1}{\sqrt{- \tilde{\tau_2}}} H_{i\mu}^{(1)} (-k_1 \tilde{\tau_2}) e^{- i k_1 \tilde{\tau_2}} \\\nonumber
      & \times \int_{-\infty}^0 d \tau_1  \frac{1}{\sqrt{- \tau_1}} H_{-i\mu}^{(2)} (-k_2 \tau_1) e^{i k_2 \tau_1} \int_{-\infty}^{\tau_1} d \tau_2  \frac{1}{\sqrt{- \tau_2}} H_{-i\mu}^{(2)} (-k_3 \tau_2) e^{i k_3 \tau_2} \bigg]                                                                                                         \\
      & \times  (2\pi)^3  \delta^3  ( {\boldsymbol k}_1+ {\boldsymbol k}_2+{\boldsymbol k}_3)+5 \,\, {\rm perm.}
\end{align}
\begin{align}\nonumber
 (3)= &
 12 c^3_2 c_3 u_{k_1} (0) u_{k_2}(0)u_{k_3}(0)\times {\rm Re} \bigg[\frac{e^{-3\pi \mu}\pi ^{3/2} \sqrt{k_1 k_2 k_3} }{16 \sqrt{2} H^4 R^3}\\\nonumber
      & \times \int_{-\infty}^{0} d \tilde{\tau_1} \sqrt{- \tilde{\tau_1}} H_{i\mu}^{(1)} (-k_1 \tilde{\tau_1}) H_{i\mu}^{(1)} (-k_2 \tilde{\tau_1}) H_{i\mu}^{(1)} (-k_3 \tilde{\tau_1})  \int_{-\infty}^0 d \tau_1  \frac{1}{\sqrt{- \tau_1}} H_{-i\mu}^{(2)} (-k_1 \tau_1) e^{i k_1 \tau_1} \\\nonumber
      & \int_{-\infty}^{\tau_1} d \tau_2 \frac{1}{\sqrt{- \tau_2}} H_{-i\mu}^{(2)} (-k_2 \tau_2) e^{i k_2 \tau_2} \int_{-\infty}^{\tau_2} d \tau_3 \frac{1}{\sqrt{- \tau_3}} H_{-i\mu}^{(2)} (-k_3 \tau_3) e^{i k_3 \tau_3} \bigg]                                                             \\
      & \times  (2\pi)^3  \delta^3  ( {\boldsymbol k}_1+ {\boldsymbol k}_2+{\boldsymbol k}_3)+5 \,\, {\rm perm.}
\end{align}
\begin{align}\nonumber
 (4)
 = &
 12 c^3_2 c_3 u^*_{k_1} (0) u_{k_2}(0)u_{k_3}(0)\times {\rm Re} \bigg[\frac{e^{-3\pi \mu}\pi ^{3/2} \sqrt{k_1 k_2 k_3}}{16 \sqrt{2} H^4 R^3}
 \times \int_{-\infty}^{0} d \tilde{\tau_1} \frac{1}{\sqrt{- \tilde{\tau_1}}} H_{i\mu}^{(1)} (-k_1 \tilde{\tau_1}) e^{-i k_1 \tilde{\tau_1}}\\\nonumber
   & \times\int_{-\infty}^0 d \tau_1 \sqrt{- \tau_1} H_{-i\mu}^{(2)} (-k_1 \tau_1)  H_{i\mu}^{(1)} (-k_2 \tau_1)  H_{i\mu}^{(1)} (-k_3 \tau_1) \int_{-\infty}^{\tau_1} d \tau_2  \frac{1}{\sqrt{- \tau_2}} H_{-i\mu}^{(2)} (-k_2 \tau_2) e^{i k_2 \tau_2} \\
   & \int_{-\infty}^{\tau_2} d \tau_3  \frac{1}{\sqrt{- \tau_3}} H_{-i\mu}^{(2)} (-k_3 \tau_3) e^{i k_3 \tau_3}\bigg]
 \times  (2\pi)^3  \delta^3  ( {\boldsymbol k}_1+ {\boldsymbol k}_2+{\boldsymbol k}_3)+5 \,\, {\rm perm.}
\end{align}
\begin{align}\nonumber
 (5)= &
 12 c^3_2 c_3 u^*_{k_1} (0) u_{k_2}(0)u_{k_3}(0)\times {\rm  Re} \bigg[\frac{e^{-3\pi \mu}\pi ^{3/2} \sqrt{k_1 k_2 k_3}}{16 \sqrt{2} H^4 R^3}
 \times \int_{-\infty}^{0} d \tilde{\tau_1} \frac{1}{\sqrt{- \tilde{\tau_1}}}H_{i\mu}^{(1)} (-k_1 \tilde{\tau_1}) e^{-i k_1 \tilde{\tau_1}} \\\nonumber
      & \times \int_{-\infty}^0 d \tau_1\frac{1}{\sqrt{- \tau_1}} H_{i\mu}^{(1)} (-k_2 \tau_1) e^{i k_2 \tau_1} \int_{-\infty}^{\tau_1} d \tau_2 \sqrt{- \tau_2} H_{-i\mu}^{(2)} (-k_1 \tau_2) H_{-i\mu}^{(2)} (-k_2 \tau_2) H_{i\mu}^{(1)} (-k_3 \tau_2) \\
      & \int_{-\infty}^{\tau_2} d \tau_3  \frac{1}{\sqrt{- \tau_3}} H_{-i\mu}^{(2)} (-k_3 \tau_3) e^{i k_3 \tau_3}\bigg]
 \times  (2\pi)^3  \delta^3  ( {\boldsymbol k}_1+ {\boldsymbol k}_2+{\boldsymbol k}_3)+5 \,\, {\rm perm.}
\end{align}
\begin{align}\nonumber
 (6)= &
 12 c^3_2 c_3 u^*_{k_1} (0) u_{k_2}(0)u_{k_3}(0)\times {\rm  Re} \bigg[\frac{e^{-3\pi \mu}\pi ^{3/2} \sqrt{k_1 k_2 k_3}}{16 \sqrt{2} H^4 R^3}
 \times \int_{-\infty}^{0} d \tilde{\tau_1} \frac{1}{\sqrt{- \tilde{\tau_1}}} H_{i\mu}^{(1)} (-k_1 \tilde{\tau_1}) e^{-i k_1 \tilde{\tau_1}}\\\nonumber
      & \times \int_{-\infty}^0 d \tau_1\frac{1}{\sqrt{- \tau_1}} H_{i\mu}^{(1)} (-k_2 \tau_1) e^{i k_2 \tau_1}
 \int_{-\infty}^{\tau_1} d \tau_2 \frac{1}{\sqrt{- \tau_2}} H_{i\mu}^{(1)} (-k_3 \tau_2) e^{i k_3 \tau_2}\\ \nonumber
      & \times \int_{-\infty}^{\tau_2} d \tau_3 \sqrt{- \tau_3} H_{-i\mu}^{(2)} (-k_1 \tau_3) H_{-i\mu}^{(2)} (-k_2 \tau_3) H_{-i\mu}^{(2)} (-k_3 \tau_3)\bigg] \\
      & \times  (2\pi)^3  \delta^3  ( {\boldsymbol k}_1+ {\boldsymbol k}_2+{\boldsymbol k}_3)+5 \,\, {\rm perm.}
\end{align}
\begin{align}\nonumber
 (7)
 = &
 12 c^3_2 c_3 u_{k_1} (0) u_{k_2}(0)u_{k_3}(0)\times {\rm Re} \bigg[\frac{e^{-3\pi \mu}\pi ^{3/2} \sqrt{k_1 k_2 k_3} }{16 \sqrt{2} H^4 R^3}\\\nonumber
   & \times \int_{-\infty}^{0} d \tau_1 \sqrt{- \tau_1} H_{i\mu}^{(1)} (-k_1 \tau_1) H_{i\mu}^{(1)} (-k_2 \tau_1) H_{i\mu}^{(1)} (-k_3 \tau_1) \int_{-\infty}^{\tau_1} d \tau_2 \frac{1}{\sqrt{- \tau_2}} H_{-i\mu}^{(2)} (-k_1 \tau_2) e^{i k_1 \tau_2} \\ \nonumber
   & \int_{-\infty}^{\tau_2} d \tau_3 \frac{1}{\sqrt{- \tau_3}} H_{-i\mu}^{(2)} (-k_2 \tau_3) e^{i k_2 \tau_3} \int_{-\infty}^{\tau_3} d \tau_4 \frac{1}{\sqrt{- \tau_4}} H_{-i\mu}^{(2)} (-k_3 \tau_4) e^{i k_3 \tau_4}\bigg]                           \\
   & \times  (2\pi)^3  \delta^3  ( {\boldsymbol k}_1+ {\boldsymbol k}_2+{\boldsymbol k}_3)+5 \,\, {\rm perm.}
\end{align}
\begin{align}\nonumber
 (8)
 = &
 12 c^3_2 c_3 u_{k_1} (0) u_{k_2}(0)u_{k_3}(0)\times {\rm Re} \bigg[\frac{e^{-3\pi \mu} \pi ^{3/2}\sqrt{k_1 k_2 k_3} }{16 \sqrt{2} H^4 R^3}\\\nonumber
   & \times \int_{-\infty}^{0} d \tau_1 \frac{1}{\sqrt{- \tau_1}} H_{i\mu}^{(1)} (-k_1 \tau_1)e^{i k_1 \tau_1}
 \int_{-\infty}^{\tau_1} d \tau_2 \sqrt{- \tau_2} H_{-i\mu}^{(2)} (-k_1 \tau_2) H_{i\mu}^{(1)} (-k_2 \tau_2) H_{i\mu}^{(1)} (-k_3 \tau_2)\\ \nonumber
   & \int_{-\infty}^{\tau_2} d \tau_3 \frac{1}{\sqrt{- \tau_3}} H_{-i\mu}^{(2)} (-k_2 \tau_3) e^{i k_2 \tau_3} \int_{-\infty}^{\tau_3} d \tau_4 \frac{1}{\sqrt{- \tau_4}} H_{-i\mu}^{(2)} (-k_3 \tau_4) e^{i k_3 \tau_4}\bigg] \\ \label{8thterms}
   & \times  (2\pi)^3  \delta^3  ( {\boldsymbol k}_1+ {\boldsymbol k}_2+{\boldsymbol k}_3)+5 \,\, {\rm perm.}
\end{align}
\begin{align}\nonumber
 (9)
 = &
 12 c^3_2 c_3 u_{k_1} (0) u_{k_2}(0)u_{k_3}(0)\times {\rm Re} \bigg[\frac{e^{-3\pi \mu}\pi ^{3/2} \sqrt{k_1 k_2 k_3} }{16 \sqrt{2} H^4 R^3}\\\nonumber
   & \times \int_{-\infty}^{0} d \tau_1 \frac{1}{\sqrt{- \tau_1}} H_{i\mu}^{(1)} (-k_1 \tau_1)e^{i k_1 \tau_1} \int_{-\infty}^{\tau_1} d \tau_2 \frac{1}{\sqrt{- \tau_2}} H_{i\mu}^{(1)} (-k_2 \tau_2)e^{i k_2 \tau_2}                                                \\\nonumber
   & \times \int_{-\infty}^{\tau_2} d \tau_3 \sqrt{- \tau_3} H_{-i\mu}^{(2)} (-k_1 \tau_3) H_{-i\mu}^{(2)} (-k_2 \tau_3) H_{i\mu}^{(1)} (-k_3 \tau_3) \int_{-\infty}^{\tau_3} d \tau_4 \frac{1}{\sqrt{- \tau_4}} H_{-i\mu}^{(2)} (-k_3 \tau_4) e^{i k_3 \tau_4}\bigg] \\
   & \times (2\pi)^3  \delta^3  ( {\boldsymbol k}_1+ {\boldsymbol k}_2+{\boldsymbol k}_3)+5 \,\, {\rm perm.}
\end{align}
\begin{align}\nonumber
 (10)
 = &
 12 c^3_2 c_3 u_{k_1} (0) u_{k_2}(0)u_{k_3}(0)\times {\rm Re} \bigg[\frac{e^{-3\pi \mu}\pi ^{3/2} \sqrt{k_1 k_2 k_3}} {16 \sqrt{2} H^4 R^3}\\\nonumber
   & \times \int_{-\infty}^{0} d \tau_1 \frac{1}{\sqrt{- \tau_1}} H_{i\mu}^{(1)} (-k_1 \tau_1)e^{i k_1 \tau_1} \int_{-\infty}^{\tau_1} d \tau_2 \frac{1}{\sqrt{- \tau_2}} H_{i\mu}^{(1)} (-k_2 \tau_2)e^{i k_2 \tau_2}                                               \\\nonumber
   & \times \int_{-\infty}^{\tau_2} d \tau_3 \frac{1}{\sqrt{- \tau_3}} H_{i\mu}^{(1)} (-k_3 \tau_3)e^{i k_3 \tau_3} \int_{-\infty}^{\tau_3} d \tau_4 \sqrt{- \tau_4} H_{-i\mu}^{(2)} (-k_1 \tau_4) H_{-i\mu}^{(2)} (-k_2 \tau_4) H_{-i\mu}^{(2)} (-k_3 \tau_4)\bigg] \\
   & \times  (2\pi)^3  \delta^3  ( {\boldsymbol k}_1+ {\boldsymbol k}_2+{\boldsymbol k}_3)+5 \,\, {\rm perm.}
\end{align}

\section{Contribution from (7), (8), (9), (10) terms}
In this section, we present the contribution from terms (7), (8), (9) and (10) separately. Adding them together gives the result  \eqref{LM_result}.

From \eqref{largemass} and \eqref{largemassint}, we can get the following useful combination
\begin{align}
   & H^{(1)}_{i\mu} (-k\tau_1) \int_{ }^{\tau_1} d\tau (-\tau)^{-1/2}  H^{(2)}_{-i\mu} (-k\tau) e^{ik\tau}   = (-2 i) \frac{e^{\pi\mu+ik\tau_1}}{\pi\mu^2} (-\tau_1)^{1/2}
\end{align}
A nice property of this expression is that many phase factors in \eqref{largemass} and \eqref{largemassint} cancel out neatly. Also note that this can be regarded as an indefinite integral. When we use it for definite integrals, the zero side and the infinity side do not contribute to the polynomially suppressed part \cite{Chen:2012ge}.  Using this expression repeatedly for the $\delta\sigma$ propagators on the $k_1$ leg and $k_2$ leg, we obtain the following contributions from the (7), (8), (9), (10) terms,
\begin{align}\nonumber
 (7) = &
 12 c^3_2 c_3 u_{k_1} (0) u_{k_2}(0)u_{k_3}(0)\times {\rm Re} \bigg[\frac{e^{-3\pi \mu}\pi ^{3/2} \sqrt{k_1 k_2 k_3} }{16 \sqrt{2} H^4 R^3}\\\nonumber
       & \times \int_{-\infty}^{0} d \tau_1 \sqrt{- \tau_1} H_{i\mu}^{(1)} (-k_1 \tau_1) H_{i\mu}^{(1)} (-k_2 \tau_1) H_{i\mu}^{(1)} (-k_3 \tau_1) \int_{-\infty}^{\tau_1} d \tau_2 \frac{1}{\sqrt{- \tau_2}} H_{-i\mu}^{(2)} (-k_1 \tau_2) e^{i k_1 \tau_2} \\ \nonumber
       & \int_{-\infty}^{\tau_1} d \tau_3 \frac{1}{\sqrt{- \tau_3}} H_{-i\mu}^{(2)} (-k_2 \tau_3) e^{i k_2 \tau_3} \int_{-\infty}^{\tau_1} d \tau_4 \frac{1}{\sqrt{- \tau_4}} H_{-i\mu}^{(2)} (-k_3 \tau_4) e^{i k_3 \tau_4}\bigg]                           \\ \nonumber
       & \times  (2\pi)^3  \delta^3  ( {\boldsymbol k}_1+ {\boldsymbol k}_2+{\boldsymbol k}_3)                                                                                                                                                               \\ \nonumber
 =     & -
 \frac{3e^{-\pi \mu}\pi}{32 H R^6}\frac{c^3_2 c_3}{k_1 k_2 k_3}\times \frac{1}{\mu^4}
 \times (2\pi)^3\delta^3 (\boldsymbol k_1+ \boldsymbol k_2 + \boldsymbol k_3  ) \\
       & \times {\rm Re}\bigg[
 \int_{-\infty}^{0} d \tau_1 (-\tau_1 )^{3/2} e^{i \tau_1  (k_1+k_2)}H_{i\mu}^{(1)} (-k_3 \tau_1) \int_{-\infty}^{\tau_1} d \tau_2 (- \tau_2)^{-1/2} H_{-i\mu}^{(2)} (-k_3 \tau_2) e^{i k_3 \tau_2} \bigg] ~.
\end{align}
The first equality follows from absorbing the 5 permutations into the integral, thus all the upper limits of the 2nd, 3rd and 4th inner integrals become $\tau_1$ \cite{Gong:2013sma}.
\begin{align}\nonumber
 (8)= &
 12 c^3_2 c_3 u_{k_1} (0) u_{k_2}(0)u_{k_3}(0)\times {\rm Re} \bigg[\frac{e^{-3\pi \mu} \pi ^{3/2}\sqrt{k_1 k_2 k_3} }{16 \sqrt{2} H^4 R^3}\\\nonumber
      & \times \int_{-\infty}^{0} d \tau_2 \sqrt{- \tau_2} H_{-i\mu}^{(2)} (-k_1 \tau_2) H_{i\mu}^{(1)} (-k_2 \tau_2) H_{i\mu}^{(1)} (-k_3 \tau_2)  \int_{\tau_2}^{0} d \tau_1 \frac{1}{\sqrt{- \tau_1}} H_{i\mu}^{(1)} (-k_1 \tau_1)e^{i k_1 \tau_1}
 \\ \nonumber
      & \int_{-\infty}^{\tau_2} d \tau_3 \frac{1}{\sqrt{- \tau_3}} H_{-i\mu}^{(2)} (-k_2 \tau_3) e^{i k_2 \tau_3} \int_{-\infty}^{\tau_2} d \tau_4 \frac{1}{\sqrt{- \tau_4}} H_{-i\mu}^{(2)} (-k_3 \tau_4) e^{i k_3 \tau_4}\bigg]                     \\ \nonumber
      & \times  (2\pi)^3  \delta^3  ( {\boldsymbol k}_1+ {\boldsymbol k}_2+{\boldsymbol k}_3)+2 \,\, {\rm perm.}                                                                                                                                      \\ \nonumber
 =    & -
 \frac{9 e^{-\pi \mu}\pi}{32 H R^6}\frac{c^3_2 c_3}{k_1 k_2 k_3}\times \frac{1}{\mu^4}
 \times (2\pi)^3\delta^3 (\boldsymbol k_1+ \boldsymbol k_2 + \boldsymbol k_3  ) \\
      & \times{\rm Re}\bigg[
 \int_{-\infty}^{0} d \tau_1 (- \tau_1)^{-1/2} H_{i\mu}^{(1)} (-k_3 \tau_1)e^{i k_1 \tau_1}
 \int_{-\infty}^{\tau_1} d \tau_2 (- \tau_2)^{3/2}e^{i(k_1+k_2)\tau_2} H_{-i\mu}^{(2)} (-k_3 \tau_2)
 \bigg] ~.
\end{align}
In obtaining the first integral above, we exchange the first and second integrals of \eqref{8thterms}, after which we perform the same trick as applied on term (7) to absorb 3 of the permutations, changing the upper limit of the last integral to $\tau_2$. Note that the method for dealing with the remaining 2 permutations is simply to write down the expression explicitly for each permutation ensuing which we integrate out the $k_1$ and the $k_2$ leg for each permutation.
\begin{align}\nonumber
 (9) = &
 12 c^3_2 c_3 u_{k_1} (0) u_{k_2}(0)u_{k_3}(0)\times {\rm Re} \bigg[\frac{e^{-3\pi \mu}\pi ^{3/2} \sqrt{k_1 k_2 k_3} }{16 \sqrt{2} H^4 R^3}\\\nonumber
       & \times \int_{-\infty}^{0} d \tau_3 \sqrt{- \tau_3} H_{-i\mu}^{(2)} (-k_1 \tau_3) H_{-i\mu}^{(2)} (-k_2 \tau_3) H_{i\mu}^{(1)} (-k_3 \tau_3) \int_{\tau_3}^{0} d \tau_1 \frac{1}{\sqrt{- \tau_1}} H_{i\mu}^{(1)} (-k_1 \tau_1)e^{i k_1 \tau_1} \\\nonumber
       & \int_{\tau_3}^{0} d \tau_2 \frac{1}{\sqrt{- \tau_2}} H_{i\mu}^{(1)} (-k_2 \tau_2)e^{i k_2 \tau_2} \int_{-\infty}^{\tau_3} d \tau_4 \frac{1}{\sqrt{- \tau_4}} H_{-i\mu}^{(2)} (-k_3 \tau_4) e^{i k_3 \tau_4}   \bigg]                          \\ \nonumber
       & \times (2\pi)^3  \delta^3  ( {\boldsymbol k}_1+ {\boldsymbol k}_2+{\boldsymbol k}_3)+2 \,\, {\rm perm.}                                                                                                                                       \\ \nonumber
 =     & -\frac{9e^{-\pi \mu}\pi}{32 H R^6}\frac{c^3_2 c_3}{k_1 k_2 k_3}\times \frac{1}{\mu^4}
 \times (2\pi)^3\delta^3 (\boldsymbol k_1+ \boldsymbol k_2 + \boldsymbol k_3  ) \\
       & \times{\rm Re}\bigg[\int_{-\infty}^{0} d \tau_3 (- \tau_3)^{3/2}e^{i(k_1+k_2)\tau_3} H_{i\mu}^{(1)} (-k_3 \tau_3)
 \int_{-\infty}^{\tau_3} d \tau_2 (- \tau_2)^{-1/2} H_{-i\mu}^{(2)} (-k_3 \tau_2) e^{i k_3 \tau_2}
 \bigg] ~.
\end{align}
\begin{align}\nonumber
 (10)= &
 12 c^3_2 c_3 u_{k_1} (0) u_{k_2}(0)u_{k_3}(0)\times {\rm Re} \bigg[\frac{e^{-3\pi \mu}\pi ^{3/2} \sqrt{k_1 k_2 k_3}} {16 \sqrt{2} H^4 R^3}\\\nonumber
       & \times \int_{-\infty}^{0} d \tau_4 \sqrt{- \tau_4} H_{-i\mu}^{(2)} (-k_1 \tau_4) H_{-i\mu}^{(2)} (-k_2 \tau_4) H_{-i\mu}^{(2)} (-k_3 \tau_4)  \int_{\tau_4}^{0} d \tau_1 \frac{1}{\sqrt{- \tau_1}} H_{i\mu}^{(1)} (-k_1 \tau_1)e^{i k_1 \tau_1} \\\nonumber
       & \int_{\tau_4}^{0} d \tau_2 \frac{1}{\sqrt{- \tau_2}} H_{i\mu}^{(1)} (-k_2 \tau_2)e^{i k_2 \tau_2} \times \int_{\tau_4}^{0} d \tau_3 \frac{1}{\sqrt{- \tau_3}} H_{i\mu}^{(1)} (-k_3 \tau_3)e^{i k_3 \tau_3}  \bigg]                              \\ \nonumber
       & \times  (2\pi)^3  \delta^3  ( {\boldsymbol k}_1+ {\boldsymbol k}_2+{\boldsymbol k}_3)                                                                                                                                                           \\ \nonumber
 =     & -
 \frac{3e^{-\pi \mu}\pi}{32 H R^6}\frac{c^3_2 c_3}{k_1 k_2 k_3}\times \frac{1}{\mu^4}
 \times (2\pi)^3\delta^3 (\boldsymbol k_1+ \boldsymbol k_2 + \boldsymbol k_3  ) \\
       & \times {\rm Re}\bigg[
 \int_{-\infty}^{0} d \tau_1 (- \tau_1)^{3/2}e^{i(k_2+k_3)\tau_1}H_{-i\mu}^{(2)} (-k_1 \tau_1)
 \int_{\tau_1}^{0} d \tau_2 (- \tau_2)^{-1/2} H_{i\mu}^{(1)} (-k_1 \tau_2)e^{i k_1 \tau_2}
 \bigg] ~.
\end{align}

\noindent (7), (8) and (9) and (10) contribute the following to the final result,
\begin{align} \nonumber
 (7) +(8) & +(9)+(10) = -\frac{e^{-\pi \mu}\pi }{2 H R^3}\frac{ \dot \theta_0^3 V'''}{k_1 k_2 k_3}\times\frac{1}{\mu^4}
 \times (2\pi)^3  \delta^3 (\boldsymbol k_1+ \boldsymbol k_2 + \boldsymbol k_3  ) \\ \nonumber
          & \times {\rm Re}\bigg[ \int_{-\infty}^0 d\tau_1 (-\tau_1)^{-1/2} H_{i\mu}^{(1)}(-k_3\tau_1) e^{ik_3\tau_1} \int_{-\infty}^{\tau_1} d\tau_2 (-\tau_2)^{3/2} H_{-i\mu} (-k_3\tau_2) e^{i\tau_2(k_1+k_2)}           \\
          & \quad \quad +\int_{-\infty}^{0} d \tau_1 (-\tau_1 )^{3/2} e^{i \tau_1  (k_1+k_2)}H_{i\mu}^{(1)} (-k_3 \tau_1) \int_{-\infty}^{\tau_1} d \tau_2 (- \tau_2)^{-1/2} H_{-i\mu}^{(2)} (-k_3 \tau_2) e^{i k_3 \tau_2}
 \bigg] ~.
\end{align}

\section{Sound Speed Corrected Bispectrum}\label{integrals}
The first integral is
\begin{align}\nonumber
   & \int_{0}^{+\infty} d\tilde z  {\tilde z}^{-1/2} H^{(1)}_{i\mu} (\tilde z)  e^{i c_s \tilde z} \int_{0}^{+\infty} d z  z^{3/2} H^{(2)}_{-i\mu} (z) e^{-i c_s \frac{ k_1+k_2}{k_3} z}                                   \\ \nonumber
 = & \int_{0}^{+\infty} d\tilde z  {\tilde z}^{-1/2} H^{(1)}_{i\mu} (\tilde z)  e^{i c_s \tilde z} \frac{i 2^{-i \mu } \Gamma (-i \mu ) }{\pi }\Gamma \left(i \mu +\frac{5}{2}\right) \left(\frac{i c_s
 \left(k_1+k_2\right)}{k_3}\right)^{-\frac{5}{2}-i \mu } \\
 + & \int_{0}^{+\infty} d\tilde z  {\tilde z}^{-1/2} H^{(1)}_{i\mu} (\tilde z)  e^{i c_s \tilde z}  \frac{2^{i \mu }  (\coth (\pi  \mu )+1) }{\Gamma (1-i \mu )}  \Gamma \left(\frac{5}{2}-i \mu \right) \left(\frac{i c_s
 \left(k_1+k_2\right)}{k_3}\right)^{-\frac{5}{2}+i \mu }
\end{align}
The second integral is
\begin{align}\nonumber
   & + \int_{0}^{+\infty} dz_1 z_1^{-1/2} H^{(1)}_{i\mu} (z_1) e^{-i c_s z_1} \int_{z_1}^{+\infty} dz_2 z_2^{3/2} H^{(2)}_{-i\mu} (z_2) e^{-i c_s \frac{k_1+k_2}{k_3} z_2}                                                                                           \\ \nonumber
   & = \int_{0}^{+\infty} dz_1 z_1^{-1/2} H^{(1)}_{i\mu} (z_1) e^{-i c_s z_1}  \bigg[\frac{i 2^{-i \mu } \Gamma (-i \mu )  }{\pi }  \Gamma\bigg(\frac{5}{2}+i\mu,\frac{i c_s z_1 (k_1+k_2)}{k_3}\bigg) \bigg( \frac{i c_s (k_1+k_2)}{k_3} \bigg)^{-\frac{5}{2}-i\mu} \\
   & +\frac{2^{i \mu }   (\coth (\pi  \mu )+1)  }{\Gamma (1-i \mu )}  \Gamma\bigg(\frac{5}{2}-i\mu,\frac{i c_s z_1 (k_1+k_2)}{k_3}\bigg) \bigg( \frac{i c_s (k_1+k_2)}{k_3} \bigg)^{-\frac{5}{2}+i\mu} \bigg]
\end{align}
The third integral is
\begin{align}\nonumber
   & + \int_{0}^{+\infty} d z_1 z_1^{3/2} H^{(1)}_{i\mu}(z_1) e^{ - i c_s  \frac{k_1 + k_2}{k_3} z_1} \int_{z_1}^{+\infty} dz_2 z_2^{-1/2} H^{(2)}_{-i\mu} (z_2) e^{-i c_s z_2}                                                                      \\ \nonumber
   & = \int_{0}^{+\infty} dz_1 z_1^{-1/2} H^{(2)}_{-i\mu} (z_1) e^{-i c_s z_1} \bigg[ \frac{2^{-i \mu } (\coth (\pi  \mu )+1) }{\Gamma (i \mu
 +1)} \bigg( \Gamma\bigg( \frac{5}{2} +i\mu \bigg) -\Gamma\bigg( \frac{5}{2} +i\mu ,\frac{i c_s z_1 (k_1+ k_2)}{k_3} \bigg)   \bigg) \bigg( \frac{i c_s (k_1+k_2)}{k_3} \bigg)^{-\frac{5}{2}-i\mu}  \\
   & -\frac{i 2^{i \mu } \Gamma (i \mu )  }{\pi } \bigg( \Gamma\bigg( \frac{5}{2} -i\mu \bigg) -\Gamma\bigg( \frac{5}{2} -i\mu ,\frac{i c_s z_1 (k_1+ k_2)}{k_3} \bigg)   \bigg) \bigg( \frac{i c_s (k_1+k_2)}{k_3} \bigg)^{-\frac{5}{2}+i\mu}\bigg]
\end{align}


\begin{thebibliography}{999}

 \bibitem{Linde:1993cn}
 A.~D.~Linde,
 ``Hybrid inflation,''
 Phys.\ Rev.\ D {\bf 49}, 748 (1994)
 [astro-ph/9307002].

\bibitem{Yamaguchi:2005qm} 
  M.~Yamaguchi and J.~Yokoyama,
  ``Density fluctuations in one-field inflation,''
  Phys.\ Rev.\ D {\bf 74}, 043523 (2006)
  [hep-ph/0512318].

 \bibitem{Chen:2009we}
 X.~Chen and Y.~Wang,
 ``Large non-Gaussianities with Intermediate Shapes from Quasi-Single Field Inflation,''
 Phys.\ Rev.\ D {\bf 81}, 063511 (2010)
 [arXiv:0909.0496 [astro-ph.CO]].

 \bibitem{Chen:2009zp}
 X.~Chen and Y.~Wang,
 ``Quasi-Single Field Inflation and Non-Gaussianities,''
 JCAP {\bf 1004}, 027 (2010)
 [arXiv:0911.3380 [hep-th]].


 \bibitem{Tolley:2009fg}
 A.~J.~Tolley and M.~Wyman,
 ``The Gelaton Scenario: Equilateral non-Gaussianity from multi-field dynamics,''
 Phys.\ Rev.\ D {\bf 81}, 043502 (2010)
 [arXiv:0910.1853 [hep-th]].
 
\bibitem{Gwyn:2014doa} 
R.~Gwyn, G.~A.~Palma, M.~Sakellariadou and S.~Sypsas,
``On degenerate models of cosmic inflation,''
JCAP {\bf 1410}, no. 10, 005 (2014)
[arXiv:1406.1947 [hep-th]].

 \bibitem{Achucarro:2010jv}
 A.~Achucarro, J.~O.~Gong, S.~Hardeman, G.~A.~Palma and S.~P.~Patil,
 ``Mass hierarchies and non-decoupling in multi-scalar field dynamics,''
 Phys.\ Rev.\ D {\bf 84}, 043502 (2011)
 [arXiv:1005.3848 [hep-th]].


 \bibitem{Cremonini:2010ua}
 S.~Cremonini, Z.~Lalak and K.~Turzynski,
 ``Strongly Coupled Perturbations in Two-Field Inflationary Models,''
 JCAP {\bf 1103}, 016 (2011)
 [arXiv:1010.3021 [hep-th]].

 \bibitem{Sasaki:1998ug}
 M.~Sasaki and T.~Tanaka,
 ``Superhorizon scale dynamics of multiscalar inflation,''
 Prog.\ Theor.\ Phys.\  {\bf 99}, 763 (1998)
 [gr-qc/9801017].

 \bibitem{Gordon:2000hv}
 C.~Gordon, D.~Wands, B.~A.~Bassett and R.~Maartens,
 ``Adiabatic and entropy perturbations from inflation,''
 Phys.\ Rev.\ D {\bf 63}, 023506 (2001)
 [astro-ph/0009131].

 \bibitem{Amendola:2001ni}
 L.~Amendola, C.~Gordon, D.~Wands and M.~Sasaki,
 ``Correlated perturbations from inflation and the cosmic microwave background,''
 Phys.\ Rev.\ Lett.\  {\bf 88}, 211302 (2002)
 [astro-ph/0107089].

 \bibitem{Peterson:2010np}
 C.~M.~Peterson and M.~Tegmark,
 ``Testing Two-Field Inflation,''
 Phys.\ Rev.\ D {\bf 83}, 023522 (2011)
 [arXiv:1005.4056 [astro-ph.CO]].


 \bibitem{Arkani-Hamed:2015bza}
 N.~Arkani-Hamed and J.~Maldacena,
 ``Cosmological Collider Physics,''
 arXiv:1503.08043 [hep-th].

 \bibitem{Lee:2016vti}
 H.~Lee, D.~Baumann and G.~L.~Pimentel,
 ``Non-Gaussianity as a Particle Detector,''
 arXiv:1607.03735 [hep-th].

 \bibitem{Chen:2016uwp}
 X.~Chen, Y.~Wang and Z.~Z.~Xianyu,
 ``Standard Model Background of the Cosmological Collider,''
 arXiv:1610.06597 [hep-th].
 
 \bibitem{Chen:2015lza}
 X.~Chen, M.~H.~Namjoo and Y.~Wang,
 ``Quantum Primordial Standard Clocks,''
 JCAP {\bf 1602}, no. 02, 013 (2016)
 [arXiv:1509.03930 [astro-ph.CO]].

 \bibitem{Chen:2016cbe}
 X.~Chen, M.~H.~Namjoo and Y.~Wang,
 ``Probing the Primordial Universe using Massive Fields,''
 arXiv:1601.06228 [hep-th].

 \bibitem{Chen:2016qce}
 X.~Chen, M.~H.~Namjoo and Y.~Wang,
 ``A Direct Probe of the Evolutionary History of the Primordial Universe,''
 Sci.\ China Phys.\ Mech.\ Astron.\  {\bf 59} (2016) no.10,  101021
 [arXiv:1608.01299 [astro-ph.CO]].


 \bibitem{Huang:2016quc}
 Q.~G.~Huang and S.~Pi,
 ``Power-law modulation of the scalar power spectrum from a heavy field with a monomial potential,''
 arXiv:1610.00115 [hep-th].

 \bibitem{Gibbons:1977mu}
 G.~W.~Gibbons and S.~W.~Hawking,
 ``Cosmological Event Horizons, Thermodynamics, and Particle Creation,''
 Phys.\ Rev.\ D {\bf 15} (1977) 2738.

 \bibitem{Mottola:1984ar}
 E.~Mottola,
 ``Particle Creation in de Sitter Space,''
 Phys.\ Rev.\ D {\bf 31} (1985) 754.

 \bibitem{Ford:1986sy}
 L.~H.~Ford,
 ``Gravitational Particle Creation and Inflation,''
 Phys.\ Rev.\ D {\bf 35} (1987) 2955.


 
 \bibitem{Meerburg:2016zdz}
 P.~D.~Meerburg, M.~Münchmeyer, J.~B.~Muñoz and X.~Chen,
 ``Prospects for Cosmological Collider Physics,''
 arXiv:1610.06559 [astro-ph.CO].
 
 
 \bibitem{Gong:2013sma}
 J.~O.~Gong, S.~Pi and M.~Sasaki,
 ``Equilateral non-Gaussianity from heavy fields,''
 JCAP {\bf 1311}, 043 (2013)
 [arXiv:1306.3691 [hep-th]].
 
 
 \bibitem{Weinberg:2008hq}
 S.~Weinberg,
 ``Effective Field Theory for Inflation,''
 Phys.\ Rev.\ D {\bf 77}, 123541 (2008)
 [arXiv:0804.4291 [hep-th]].
 
 
 \bibitem{Achucarro:2012sm}
 A.~Achucarro, J.~O.~Gong, S.~Hardeman, G.~A.~Palma and S.~P.~Patil,
 ``Effective theories of single field inflation when heavy fields matter,''
 JHEP {\bf 1205}, 066 (2012)
 [arXiv:1201.6342 [hep-th]].

 \bibitem{Achucarro:2010da}
 A.~Achucarro, J.~O.~Gong, S.~Hardeman, G.~A.~Palma and S.~P.~Patil,
 ``Features of heavy physics in the CMB power spectrum,''
 JCAP {\bf 1101}, 030 (2011)
 [arXiv:1010.3693 [hep-ph]].

 \bibitem{Chen:2012ge}
 X.~Chen and Y.~Wang,
 ``Quasi-Single Field Inflation with Large Mass,''
 JCAP {\bf 1209}, 021 (2012)
 [arXiv:1205.0160 [hep-th]].

 \bibitem{Pi:2012gf}
 S.~Pi and M.~Sasaki,
 ``Curvature Perturbation Spectrum in Two-field Inflation with a Turning Trajectory,''
 JCAP {\bf 1210}, 051 (2012)
 [arXiv:1205.0161 [hep-th]].

 
 \bibitem{Mirbabayi:2015hva}
 M.~Mirbabayi and M.~Simonović,
 ``Effective Theory of Squeezed Correlation Functions,''
 JCAP {\bf 1603}, no. 03, 056 (2016)
 [arXiv:1507.04755 [hep-th]].
 
 

 \bibitem{Noumi:2012vr}
 T.~Noumi, M.~Yamaguchi and D.~Yokoyama,
 ``Effective field theory approach to Quasi-Single field inflation and effects of heavy fields,''
 JHEP {\bf 1306}, 051 (2013)
 [arXiv:1211.1624 [hep-th]].


\bibitem{Gwyn:2012mw}
 R.~Gwyn, G.~A.~Palma, M.~Sakellariadou and S.~Sypsas,
 ``Effective field theory of weakly coupled inflationary models,''
 JCAP {\bf 1304}, 004 (2013)
 doi:10.1088/1475-7516/2013/04/004
 [arXiv:1210.3020 [hep-th]].
 
 
 \bibitem{Baumann:2011su}
 D.~Baumann and D.~Green,
 ``Equilateral Non-Gaussianity and New Physics on the Horizon,''
 JCAP {\bf 1109}, 014 (2011)
 [arXiv:1102.5343 [hep-th]].

 \bibitem{An:2017hlx}
 H.~An, M.~McAneny, A.~K.~Ridgway and M.~B.~Wise,
 ``Quasi Single Field Inflation in the non-perturbative regime,''
 arXiv:1706.09971 [hep-ph].

 \bibitem{Tong:2017iat}
 X.~Tong, Y.~Wang and S.~Zhou,
 ``On the Effective Field Theory for Quasi-Single Field Inflation,''
 arXiv:1708.01709 [astro-ph.CO].
 
 
 \bibitem{Chen:2017ryl}
 X.~Chen, Y.~Wang and Z.~Z.~Xianyu,
 ``Schwinger-Keldysh Diagrammatics for Primordial Perturbations,''
 arXiv:1703.10166 [hep-th].
 
 

 \bibitem{Chen:2015dga}
 X.~Chen, M.~H.~Namjoo and Y.~Wang,
 ``On the equation-of-motion versus in-in approach in cosmological perturbation theory,''
 JCAP {\bf 1601}, no. 01, 022 (2016)
 [arXiv:1505.03955 [astro-ph.CO]].

\end{thebibliography}
\end{document}